\documentclass[twocolumn]{aastex63}

\usepackage[super]{nth}

\usepackage[titletoc,title]{appendix}

\shorttitle{LIN 358 and SMC N73}
\shortauthors{Washington et al.}

\graphicspath{{./}{figures/}}

\newcommand{\UVA}{Department of Astronomy, University of Virginia, Charlottesville, VA 22904, USA}
\newcommand{\IAC}{Instituto de Astrof\'isica de Canarias, E-38205 La Laguna, Tenerife, Spain}
\newcommand{\ULL}{Universidad de La Laguna, Dpto. Astrof\'isica, E-38206 La Laguna, Tenerife, Spain}

\newcommand{\MSU}{Department of Physics, Montana State University, Bozeman, MT 59717, USA}
\newcommand{\NSFOIR}{NSF OIR Lab, Tucson, AZ 85719, USA}
\newcommand{\Steward}{Steward Observatory, Department of Astronomy, University of Arizona, Tucson, AZ 85721, USA}
\newcommand{\Brasil}{Observat\'orio Nacional, Rio de Janeiro, Brazil}
\newcommand{\Vanderbilt}{Department of Physics and Astronomy, Vanderbilt University, Nashville, TN 37235, USA}

\begin{document}

\title{Symbiotic Stars in the APOGEE Survey: The Case of LIN 358 and SMC N73 (LIN 445a)}

\correspondingauthor{Jasmin Washington}
\email{washingtonj@email.arizona.edu}

\author[0000-0002-7046-0470]{Jasmin E.\ Washington}
\affiliation{\Steward}
\affiliation{\UVA}

\author[0000-0002-7871-085X]{Hannah M.\ Lewis}
\affiliation{\UVA}

\author[0000-0001-5261-4336]{Borja Anguiano}
\affiliation{\UVA}

\author{Steven R.\ Majewski}
\affiliation{\UVA}

\author{S.\ Drew Chojnowski}
\affiliation{\MSU}

\author[0000-0002-0134-2024]{Verne V.\ Smith}
\affil{\NSFOIR}

\author[0000-0002-3481-9052]{Keivan G.\ Stassun}
\affil{\Vanderbilt}

\author[0000-0002-0084-572X]{Carlos Allende Prieto}
\affiliation{\IAC}
\affiliation{\ULL}

\author[0000-0001-6476-0576]{Katia Cunha}
\affiliation{\Steward}
\affiliation{\Brasil}

\author{David L.\ Nidever}
\affiliation{\MSU}

\author[0000-0002-1693-2721]{D.\ A.\ Garc\'ia-Hern\'andez}
\affiliation{\IAC}
\affiliation{\ULL}

\author{Kaike Pan}
\affiliation{Apache Point Observatory and New Mexico State University, P.O. Box 59, Sunspot, NM, 88349-0059, USA}

\begin{abstract}

LIN 358 and SMC N73 are two symbiotic binaries in the halo of the Small Magellanic Cloud, each composed of a hot white dwarf accreting from a cool giant companion. In this work, we characterize 
these systems using a combination of SED-fitting 
to the extant photometric data spanning a broad wavelength range (X-ray/ultraviolet to near-infrared), detailed analysis of the APOGEE spectra for the giant stars, and orbit fitting to high quality radial velocities from the APOGEE database.
Using the calculated Roche lobe radius for the giant component and the mass ratio for each system, it is found that LIN 358 is likely undergoing mass transfer via wind Roche lobe overflow while the accretion mechanism for SMC N73 remains uncertain.
This work presents the first orbital characterization for both of these systems (yielding periods of $>$270 and $>$980 days, respectively, for SMC N73 and LIN 358) and the first global SED fitting for SMC N73. In addition, variability was identified in APOGEE spectra of LIN 358 spanning 17 epochs over two years that may point to a time variable accretion rate as the product of an eccentric orbit.

\end{abstract}


\section{Introduction} \label{sec:intro}

Symbiotic stars (hereafter SySts) are interacting binary systems composed of a compact stellar remnant, 
typically white dwarfs (WDs) but also neutron stars,
accreting from a cool, giant companion (e.g., \citealt{Muerset1999}), most likely via a stellar wind (e.g., \citealt{Muerset1999}), or Roche lobe overflow (RLOF, e.g.,
\citealt{mikolajewska2003b,Munari2019}. 
SySts are further classified as S-type, for those where the mass donor is a normal giant star, and D-type, for systems where the donor is a Mira variable surrounded by a warm dust envelope \citep{webster1975}.
SySts are useful tools for the study of the evolution and effects of mass loss caused by stellar interactions in detached and semi-detached binaries. 
Each component of the SySt can show intrinsic variability with differing timescales, making them particularly effective astrophysical laboratories for the study of stellar binary evolution \citep[e.g.,][]{Podsiadlowski2008}. 

Unfortunately, however, the number of confirmed SySts is 
lower than the predicted number based on population synthesis of the observed number of white dwarfs with red giant (RG) or asymptotic giant companions \citep{Lu2006}, and
their statistics are complicated
by the fact that 
their spectra, at least for D-types, can be confused with those from other objects, such as planetary nebulae (PNe) or dense H II regions \citep{belczynski2000}.
With more careful analysis, however, it has been shown that D-type SySts can, in fact, be distinguished from PNe by the strength of forbidden lines (O III and N III) and various He I lines in the optical region of the spectrum \citep{gomezmoreno1995,Ilkiewicz2017}.
Meanwhile, S-type SySts are typically identified by the presence of photospheric absorption features (TiO, VO, C$_2$, CN) in the observed spectrum of the cool companion, in addition to presence of strong emission lines 
from He II 
and H$\alpha$, as well as 
from high-excitation ions \citep[Fe VII, O VI; e.g.,][]{belczynski2000}.

Using these (and other) criteria, a statistically large enough number of Milky Way SySts have been found and further characterized with follow-up observations such that global properties of the population are becoming well understood.  For example, orbital periods for Galactic systems tend to be between 1 and 3\,years, but can be significantly longer.  Moreover, based on the orbital parameters for $\sim$30 SySts,
Galactic symbiotic stars tend to have nearly circular orbits, though significant eccentricities ($e \gtrsim 0.1$) have been found for systems with periods longer than 1000\,days \citep{mikolajewski2003}.

Given the presence of an overly extended, cool evolved companion, SySts are among the intrinsically brightest variable stars and can be easily detected in nearby galaxies \citep[e.g.,][]{Mikolajewska2014},
and in particular in the Magellanic Clouds \citep{Angeloni2014}.  Despite this fortunate circumstance, very little is known 
about the properties of SySts in these or other galaxies. 
While a growing number of confirmed extragalactic SySts are being identified---including ten in the Large Magellanic Cloud (LMC) and twelve in the Small Magellanic Cloud (SMC; \citealt{Merc2019}, and references therein)---and such systems are accessible to current technology for the follow-up observations needed for orbital characterization, 
to date, full Keplerian orbital parameters have been derived for only one extragalactic SySt---Draco C1 \citep{Lewis2020}. The definitive binary kinematics measured in the latter work benefited from serendipitous and repeated targeting of this source by the Apache Point Observatory Galactic Evolution Experiment (APOGEE) survey \citep{Maj2017} throughout the system's 3.3 year orbital period, in the course of APOGEE's exploration of bright stars in Milky Way satellite galaxies.

\begin{figure}[h!]
\centering
\includegraphics[width=\columnwidth]{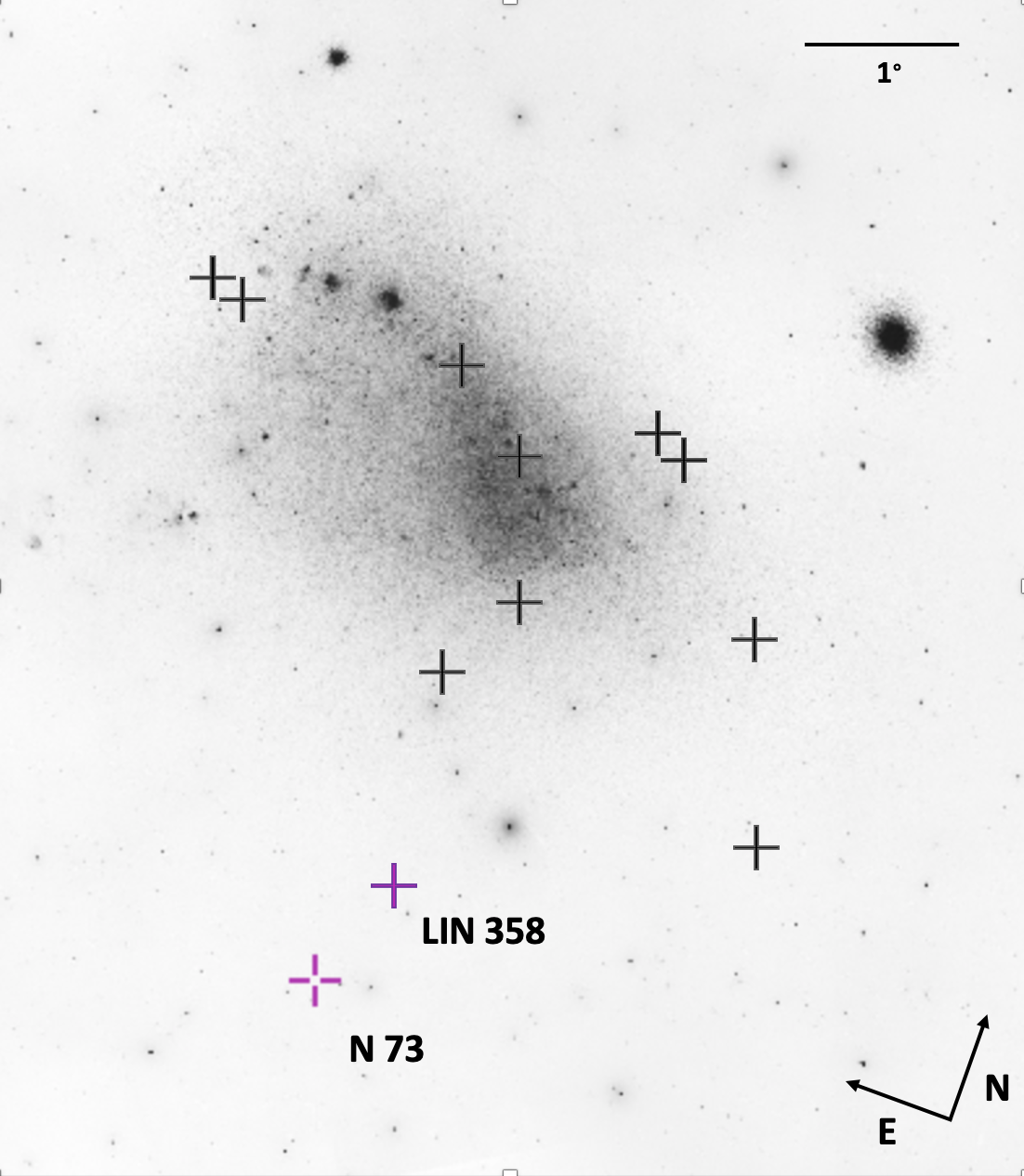}
\caption{A Digitized Sky Survey (DSS) image of the Small Magellanic Cloud \citep{DSS} showing the position of known SMC symbiotic binaries \citep{Merc2019} as crosses.  The two SySts that are the subject of the present study (shown as colored crosses) happen to be the two of twelve known that are at the largest angular separation from the SMC.}
\label{fig:SMC}
\end{figure}

Here we continue these efforts 
to analyze detailed time-domain data for the characterization of extragalactic SySt architectures using spectroscopy from both APOGEE and the succeeding APOGEE-2 survey (Majewski et al. 2021),
focusing on two systems in the SMC.
SMC N73 (LIN 445a) and LIN 358 are two known symbiotic binaries in the extreme outskirts of this galaxy.  In fact, of the known SMC SySts, these two distinguish themselves as being at largest projected radii from the SMC center (2.5$^{\circ}$ for LIN 358 and 3.1$^{\circ}$ for SMC N73; Figure \ref{fig:SMC}), which is remarkably remote, 
as previously noted by, e.g., \citet{Haberl2000}.
Both are classified as 
S-type binaries based on the near-infrared color temperature of $\sim$3000--4000\,K, which indicates the presence of a G, K, or M spectral-type giant \citep{Skopal2005, Akras2019}.
We summarize other relevant information about these two systems below.

\subsection{SMC N73: Summary of Previous Relevant Work}

SMC N73 first appears in literature in the 1950's 
as the emission nebula $\mathrm{LH\alpha}$ 115-N73 and as one of the outlying SMC systems in the catalogue by \citet{SMC1956}. SMC N73 (as 115-N73) was noted to have somewhat faint $\mathrm{H\alpha}$ emission intensity relative to the background continuum.
\citet{Lindsay61}, who labeled the system LIN 445a, was the first to question whether this system was actually a PNe, but did still list it as a probable PNe. \citet{Sanduleak1981} later classified it as a possible VV Cephei-type binary star (i.e., a system with a composite spectral energy distribution suggesting an evolved late type star with mass transfer onto a hot early type star made evident by H$\alpha$ emission),
and noted the presence of a strong ultraviolet (UV) continuum as well as TiO absorption. Based on the presence of both MgH and TiO absorption in the optical spectrum of SMC N73, the giant star has been classified as an oxygen-rich mid-K spectral type \citep{Morgan1988,Muerset96,Muerset1999}.

\citet{Walker1983} 
established 
SMC N73's status as a symbiotic star based on finding strong He II $\lambda4686$ emission at
0.59 the intensity of  $\mathrm{H\beta}$.
\citet{Morgan1992} then addressed the spectral variability of N73, stating that he had obtained a spectrum (unpublished) 
showing a He II to $\mathrm{H\beta}$ ratio of $\sim$0.9, i.e., larger than Walker's published value.
Other authors (e.g., \citealt{Morgan1992}, \citealt{Winckel93}, and \citealt{Akras2019}) affirmed the SySt status of SMC N73  based on observations of He I, O VI, H$\alpha$ in its spectrum, and,
based on ROSAT X-ray data \citep[soft X-rays at $\sim$0.28--1\,keV; ][]{trumper1984}, \citet{Bickert1996} identified the compact companion of N73 as a WD.
From the continuum UV, \citet{Muerset96} derived the radius of that WD to be $0.10$\,R$_\odot$ and its  temperature to be 130,000\,K.

While SMC N73 has clearly received ample prior spectroscopic attention at optical wavelengths, a number of key parameters of the components of the system remain relatively poorly established, 
either due to inadequate spectral resolution or relatively meager coverage in other regions of the spectrum.
For example, the previous measurements for the WD temperature in SMC N73 range from 130,000 to 200,000\,K 
\citep{Morgan1992,Muerset96}, whereas estimates of the effective temperature and radius of the giant in SMC N73 are fairly consistent among one another, with $T_\mathrm{eff} \sim 3850$\,K and $R_\mathrm{RG} \sim 150$\,R$_{\odot}$ \citep{Muerset96,Akras2019}.
Meanwhile, the orbital properties of the binary are completely unknown.  In this paper we exploit both broadband photometric measurements and the high resolution, multi-epoch infrared (IR) APOGEE spectra to improve not only the stellar atmospheric characterization of each of the components of SMC N73, but their binary orbital parameters as well.
In addition, we show that SMC N73 has not exhibited any spectral variability in the APOGEE observations, which, however, only cover a relatively short timescale---30 days---compared to the typical orbital period of symbiotic stars.

\subsection{LIN 358: Summary of Previous Relevant Work}
\label{sec:introLIN358}

LIN 358 was discovered by \citet{Lindsay61}, where the system was listed as a probable PN with moderately strong H$\alpha$ emission.
LIN 358's status as a PNe was later debated, with subsequent spectroscopic analyses recording---or not---the presence of other emission features \citep{Sanduleak1981,Walker1983,Morgan1992,Winckel93,Muerset96}.
So too has the spectral variability of LIN 358 been controversial, particularly for H$\alpha$, with \citet{Morgan1992} claiming no variability on decade timescales but \citet{Sanduleak1981} suggesting 
variability on the timescale of years. 
However, the ASAS-SN photometric catalog \citep{shappee2014,Jayasinghe2020} identifies LIN 358 as a red, irregular variable star with a $V$-band variability amplitude of 0.44\,mag. Irregular variable here means non-periodic variability which cannot be explained by, e.g., a transit or pulsational variability.

\citet{Walker1983} first pointed out that LIN 358 may be a SySt, based on a variety of spectroscopic features he found: On the blue end of the spectrum, very strong He II emission lines at 0.54 the intensity of the $\mathrm{H\beta}$ emission, a weak continuum, and no nebular emission lines, while, in the red, a smooth continuum (see also \citealt{Morgan1988}) with no visible absorption bands, but with enhanced He I at 0.27 the intensity of $\mathrm{H\alpha}$ emission.
The suggestion that LIN 358 is a SySt was supported by \citet{Vogel1995}, who used the Hubble Space Telescope (HST) to obtain an UV spectrum of LIN 358
in which they noted an overabundance of nitrogen emission, possibly from a thermonuclear event; the latter, if true, would make LIN 358 a nova, an interpretation consistent with the observed spectroscopic and photometric variability.
\citet{Haberl2000} subsequently detected X-ray emission from LIN 358 and classified the system as a supersoft source (SSS).  The supersoft X-ray emission observed from LIN 358 
is understood to be from the steady nuclear burning of hydrogen accreted from the bloated, cooler primary onto the surface of its WD companion, and the photosphere of the WD produces photons with $h\nu\approx0.2$ keV \citep{Muerset1997,Skopal2015}. 
In addition, the low $N_\mathrm{H}$ measured by e.g., \citet{Kahabka2006} and \citet{Orio2007} are indicative of low density gas near the hot WD (and at a large distances from the red giant; \citealt{Saeedi2018}).

By now, it is
firmly established that the LIN 358 binary system contains a cool, K-type giant \citep{Sanduleak1981,Muerset96,Muerset1999}, with a temperature in the range 3814--4000 K \citep{Muerset96,Akras2019}.  
By a detailed multiwavelength spectral energy distribution (SED) fit (see Section \ref{subsec:sed358} below),
\citet{Skopal2015} classified the giant as a K5 Ib supergiant.
Meanwhile, the companion star is a white dwarf with a temperature in the range 140,000 to 250,000 K and a radius between $0.09$ and $0.23R_\odot$  \citep{Morgan1992,Vogel1995,Muerset96,Kahabka2006,Orio2007,Skopal2015}.

From the emission line spectra of LIN 358, \cite{Kuuttila2020} also derive the WD bolometric luminosity $L = 1.02 \times 10^{38}$\,erg\,s$^{-1}$ (which agrees with that derived by \citealt{Skopal2015}) and the mass-loss rate of 
the giant star to be 
$1.2 \times 10^{-6}$\,M$_\odot$\,yr$^{-1}$. Assuming that the orbital separation for the system is too large for standard Roche lobe overflow (RLOF), as is the case for most symbiotic binaries \citep{mikolajewski2003,mikolajewska2003b,Mikolajewska2007,Munari2019},
the WD must be accreting via wind Roche lobe overflow \citep[WRLOF; ][]{mohamed2007,mohamed2012}, and therefore it is possible to estimate the orbital separation of LIN 358 using the ratio of the wind acceleration radius and the Roche lobe radius. Adopting a mass of 5\,M$_\odot$ for the giant star, and a mass of 1\,M$_\odot$ for the WD (given the X-ray spectrum presented by \citealt{Orio2007}), \cite{Kuuttila2020} derive a semi-major axis $a = 3.7$\,AU, which corresponds to an orbital period of 2.9\,yrs. However, this derivation assumes that the accretion efficiency is at a maximum over the entire orbit \citep[implying that the orbit is circular;][]{abate2013}, even though SySts with periods longer than 1000 days ($\sim$2.7\,yrs) have been found to have significant eccentricities \citep{mikolajewski2003}. 

Unfortunately, without a kinematical derivation of the orbital parameters for LIN 358, it has not been possible to be certain of the detailed physical mechanisms at play in the LIN 358 system. This includes definitive knowledge of the period of the system, which drives the timescales of variability, as well as  the masses, radii and minimum separation of the two stars, which, in turn, determine the accretion rates.

Fortunately, however, the APOGEE-2 survey has targeted LIN 358 repeatedly over the course of $\sim$2\,years, which provides the opportunity not only to clarify the properties of the two stars in the binary, but
refine the orbit of this system and uncover substantial spectroscopic variability over the course of these observations.  Together, these new observations help to fill in a more detailed portrait of this SySt, and the physics of its interacting components.

\subsection{Layout of this Study}

The layout of this paper is as follows:
In Section \ref{sec:data}, we discuss the sources of data used in our analysis, along with the Monte Carlo sampler \textit{The Joker}, which is used to constrain the orbits of the binaries.
Section \ref{sec:n73} briefly describes the SED-fitting process employed to find the radius of both stars and the mass of the giant in SMC N73, along with the results of the orbital analysis. 
Section \ref{sec:lin358} compares the results of the \citet{Skopal2015} analysis of the SED for LIN 358 to previous literature values and APOGEE measurements of the giant in the binary, and presents the results of the orbital fitting.
Section \ref{sec:discussion} examines the most likely mass accretion mechanisms for the two systems based on the Roche lobe radius and mass ratio.
Finally, Section \ref{sec:summary} discusses the derived characteristics of the systems and summarizes the key findings of this work.

The present study follows on, and in some ways parallels, our previous work on the Draco C1 system in the Draco dwarf spheroidal satellite of the Milky Way \citep{Lewis2020}.  Our overall exploration of symbiotic stars in Milky Way satellite galaxies derives from a larger, more systematic survey of binary systems with confirmed or suspected white dwarf components in the APOGEE survey \citep{Anguiano20}. The analyses here and in \citet{Lewis2020} show the usefulness of the APOGEE survey in characterizing these compelling and astrophysically important star systems.

\section{Data} \label{sec:data}
The second phase of the Apache Point Observatory Galactic Evolution Experiment (APOGEE-2; Majewski et al., in prep.), 
part of the Sloan Digital Sky Survey IV (SDSS-IV; \citealt{blanton2017})
employs twin spectrographs \citep{Wilson_2019} on the SDSS 2.5-m at Apache Point Observatory \citep{Gunn06} in New Mexico, and the du Pont 2.5-m telescope at Las Campanas Observatory (LCO), Chile.  Combined with the results from APOGEE \citep{Maj2017} in SDSS-III \citep{2011AJ....142...72E}, the now near-complete total APOGEE database provides high-resolution, multi-epoch spectra in the $H$-band of about half a million stars sampling all Milky Way stellar populations, as well as those of nearby Local Group dwarf galaxies \citep{ahumada2020,jonsson2020}. In particular, the Southern Hemisphere access afforded by the new spectrograph at LCO enables substantial APOGEE-2 coverage of the Magellanic Clouds
\citep[e.g.,][]{Nidever20}.

Because of the faint magnitudes of even red giant (RG) stars in the Clouds, to achieve the nominal APOGEE signal-to-noise ratio $\mathrm{S/N} > 100$ criterion for measuring chemical abundances, the survey requires a minimum of nine $\sim$1 hour APOGEE-2 visits for each LMC field, and twelve such visits for each SMC field.  These visits are spread out over many nights spanning months and even years.  While APOGEE stellar atmospheric parameters and chemical abundances are measured from co-added exposures for each target, a quality RV---at $\sim$100--200\,m\,s$^{-1}$ precision---can be measured from each individual visit spectrum \citep{Nidever15}.
These time series data provide the opportunity to look for RV variations 
indicating
binary companions \citep[e.g.,][]{Troup16,Badenes18,Price-Whelan18,JokerVAC,Mazzola20}.

In this paper, we exploit the high-resolution APOGEE spectra to derive stellar parameters, as well as the RV time series data to refine our knowledge of the stellar constituents and orbits of the SMC symbiotic binaries LIN 358 and SMC N73. 
While the RG component of LIN 358 has been well-characterized previously \citep{Skopal2015}, 
we make use of the inferred effective temperature for the RG primaries of these systems based on the APOGEE Stellar Parameters and Chemical Abundance Pipeline \citep[ASPCAP;][]{aspcap}. However, while the effective temperature is one of the most robust stellar parameters derived by ASPCAP, it has been shown that the automated ASPCAP pipeline does not work well for the most luminous stars (i.e., those with lowest surface gravity; \citealt{Schultheis2020}). For this reason, we assume the ASPCAP-derived and \cite{Skopal2015}-derived $T_\mathrm{eff}$ for the RG components of SMC N73 and LIN 358, respectively, 
but determine the spectroscopic $\log{g}$ and iron [Fe/H] abundances manually, using the Ti I / Ti II lines measured from the combined APOGEE spectra \citep{smith2021}.
Because the ASPCAP best-fit places SMC N73 at a synthetic spectral library grid edge no calibrated parameters were provided for the system, so we used the raw value for $T_{\rm{eff}}$, the only RG parameter not obtained with a boutique analysis of the APOGEE spectrum.
The assumed and derived parameters, including $T_\mathrm{eff}$, $\log{(g)}$, and [Fe/H], for both SMC N73 and LIN 358 are reported in Table \ref{tab:params}. 
In addition to deriving Fe and Ti abundances that result from the use of Ti I/Ti II to determine surface gravity, the Ce II lines that fall within the APOGEE window \citep{cunha2017} were analyzed to estimate an s-process abundance in each SMC N73 and LIN 358.  Cerium is overabundant, relative to Fe, in both stars, with [Ce/Fe]$\sim$+1.0\,dex in LIN 358 and [Ce/Fe]$\sim$+0.5\,dex in SMC N73.  The enhancement of s-process elements, such as Ce, is not unusual in metal-poor S-type symbiotic stars, as was found by \cite{smith1996} and \cite{smith1997}.

Further, we apply \textit{The Joker} \citep{Joker2017} to the RV time-series to provide constraints on Keplerian parameters for these two symbiotic systems, including the orbital period $P$, eccentricity $e$, velocity semi-amplitude $K$, and the system barycenter velocity $v_0$.
\citet{JokerVAC} details the six nonlinear and two linear parameters sampled over by \textit{The Joker}, along with their prior probability distribution functions (pdfs). In this work, the only change is in the assumed prior pdf for $v_0$. Instead of adopting the prior assumed by \citet{JokerVAC} for Milky Way stars, we assume a normal distribution 
based on the spectroscopic study by \citet{Harris2006}, which found that RG stars in the SMC have a mean velocity of $145.6\pm0.6$\,km\,s$^{-1}$ and a velocity dispersion of $27.6\pm0.5$\,km\,s$^{-1}$.
We generate a cache of $2^{24}$ dense prior samples, and rejection sample to produce
the requested number of prior samples ($M_\mathrm{min}$); in this work, the user-defined number of requested samples is $M_\mathrm{min} = 512$. In the rejection sampling step, we also require that the minimum companion mass $M_\mathrm{WD,\,min}$---derived by sampling over the reported uncertainties on the primary mass, assuming a Gaussian noise distribution on the primary mass---have a mass greater than 0.1\,M$_\odot$ (a mass smaller than the least-massive known WD, \citealt{Kilic2007}) but less than the Chandrasekhar mass (1.4\,M$_\odot$).
If, following iterative rejection sampling by \textit{The Joker}, fewer than the requested number of samples are returned but the samples are unimodal, we initialize Markov chain Monte Carlo (MCMC) to continue generating samples; otherwise, we continue sampling with \textit{The Joker} until the minimum number of samples is reached.

To fully classify both the hot and cool components of these symbiotic systems, we depend on interpretation of the system spectral energy distribution (SED), derived from broadband photometric measurements spanning a wide range of wavelengths. For SMC N73, the SED wavelengths range from the \textit{GALEX} far-ultraviolet (at $\sim$150\,nm) to the \textit{WISE} mid-IR (W3 filter at $\sim$10,000\,nm). For LIN 358, the SED ranges from the soft X-ray to near-IR $\sim$3.1--2200\,nm \citep{Skopal2015}. For LIN 358 we make use of the thorough SED-analysis presented by \citet{Skopal2015} (including the derived radius for the RG component), for which we verify the temperature for the cooler star derived by ASPCAP from the high-resolution APOGEE spectra (Section \ref{sec:lin358}).  Meanwhile, for N73, we assume the stellar parameters derived from the APOGEE spectrum, and use our own SED-fitting pipeline \citep[following the methods of][]{Stassun2017}, summarized in Section \ref{sec:sed73}, to constrain the radius of the giant and WD.

By assuming these precise stellar-parameters for the giant component of SMC N73, the number of free parameters in the SED fitting process is reduced, producing a more accurate estimate for the radii of both binary components and for the WD temperature.
Observational data for SMC N73 was obtained from the Strasbourg astronomical Data Center (CDS) portal.

Throughout this paper, we assume a distance modulus for the SMC of $\mu = 19.01 \pm 0.08$ \citep[Cepheid distance,][]{distmod},
which corresponds to a distance of $d = 63.4 \pm 2.3$\,kpc.

\section{Analysis of SMC N73} \label{sec:n73}
\subsection{SED Fitting} \label{sec:sed73}

Despite the existence of panchromatic photometry for N73 from the ultraviolet to the infrared, to date no global SED fitting has been performed on this source over that full range of available wavelengths to estimate the properties of both the hot and cool components of this SySt.  \cite{Akras2019} have used the combined {\it Gaia}, 2MASS and WISE data to constrain the photometric and thermal properties of the red giant and associated dust (see below), but address neither the properties of the hot component nor the mass and radius of the giant. This we do here using two independent procedures: (1) following the methods laid out by \citet{Stassun2016} and \citet{Stassun2017}, and (2) fitting 
a Kurucz RG stellar model \citep{Kurucz2013} to the optical-IR wavelength SED, followed by fitting to the residual flux across UV through IR wavelengths
a simple blackbody representing the 
WD.\footnote{This alternative code, WHACCKEY PAR (WHite dwarf And Companion Characterization of KEY PARameters), is presented in \citet{Washington2020}.} Both methods assume an extinction $A_V = 0.17^{+0.00}_{-0.02}$, the full extinction for the line of sight from the Galactic dust maps of \citet{Schlegel1998}, and adopt the same template SEDs for the two components (i.e., a \citealt{Kurucz2013} RG plus hot blackbody),
so it is not surprising that they yield nearly identical results. 
The primary differences between the methods are that (a) method (1) interpolates within the model grid to find the appropriate atmosphere for each star, whereas method (2) selects the nearest, best-matching models, and (b) in method (1) the separate fits are summed and scaled by each stars' surface area and to obtain the final best fit the extinction $A_V$ and overall normalization are varied to obtain the minimized $\chi^2$ \citep{Stassun2016}, whereas in method (2) there is no $\chi^2$ test performed on the sum of the two models.
In the end, we adopt the results from method (1) in this work (because the methodology presented by \citealt{Stassun2016} has been more thoroughly vetted in the literature than have those of the second algorithm). Method (2) has provided the basis for our ongoing work on other WD-binary systems (e.g., Anguiano et al., in prep.) and serves as an independent check for (1).

\begin{figure}[ht!]
    \centering
    \includegraphics[trim={3.5cm 2.5cm 3cm 3cm}, clip, width=\columnwidth]{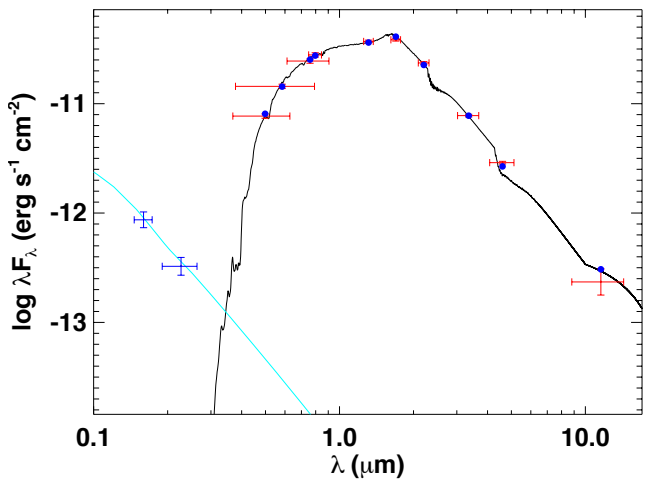}
    \caption{SED fit to combined Gaia, GALEX, 2MASS and WISE photometry for SMC N73. Red crosses represent observed broadband fluxes from Gaia $G_{\rm BP}$ at 505 nm to WISE W3 at 12 $\mu$m, and blue
    crosses show the GALEX NUV and FUV fluxes. The black curve is the Kurucz synthetic spectrum, fit to the red giant primary, and the cyan curve is the blackbody representing the white dwarf. Blue points mark the modeled flux at the effective wavelength of the filter passbands (i.e., the product of the transmission function of the filter multiplied by the flux of the star as a function of wavelength) and so can differ slightly from the Kurucz spectrum at the same  wavelength.}
    \label{fig:sed73}
\end{figure}

For the giant star, we can exploit the additional information provided by the APOGEE spectrum, and therefore assume the stellar parameters derived from the APOGEE spectra (presented in Table \ref{tab:params})---namely $T_\mathrm{eff} = 3590 \pm 50$\,K (as derived by ASPCAP), $\log{(g)} = 0.10 \pm 0.10$\,where $g$ is in units of cm\,s$^{-2}$, and $[\mathrm{Fe/H}] = -0.75 \pm 0.12$ (the latter two parameters derived by the boutique analysis discussed in Section \ref{sec:data})---for the Kurucz model and fit the atmosphere model to the flux measurement, minimizing the $\chi^2$ by varying only a single scaling factor, $\left( R_\mathrm{RG}/d \right)^2$, where we assume a distance of $d = 63.4 \pm 2.3$\,kpc  \citep{distmod}. The resulting best-fit model, shown in Figure \ref{fig:sed73} (black), has a reduced $\chi^2 = 25.9$ and shows good agreement with the observed flux measurements from the Gaia $G_{\rm BP}$ band (effective wavelength 505.0\,nm) to WISE W3 (effective wavelength 12.1\,$\mu$m), and yields the radius of the giant component as $R_\mathrm{RG} = 208 \pm 16$\,R$_\odot$. In turn, we use this radius to calculate the mass of the giant using the spectroscopically derived $\log{(g)}$, and find 
$M_{\rm RG} = 2.00 \pm 0.55$\,M$_\odot$.

For the WD component, which contributes excess flux at the bluest wavelengths, we fit a simple blackbody to the GALEX NUV and FUV bands (no X-ray data are available for SMC N73, unlike the case for LIN 358). The blackbody is fit by varying only the WD temperature and a scaling factor $\left( R_\mathrm{WD}/d \right)^2$, where we again assume the distance derived by \citet{distmod}. The model (cyan in Figure \ref{fig:sed73})
is only fit to the two GALEX data points, making a precise temperature estimate impossible, as the Rayleigh-Jeans tail of the blackbody has nearly identical profiles between the GALEX NUV and FUV fluxes at temperatures $>10^5$\,K, a typical temperature for most WDs in SySts. At such a high temperature, we can obtain the same SED fit by many different combinations of temperature and radius scalings.  Given this ambiguity, 
we can test various hypotheses to constrain the likely WD temperature.  If we assume a WD radius of 0.1\,R$_\odot$ ($\sim$10.9\,R$_\oplus$, which is a typical radius for WDs in SySts during quiescence; \citealt{Muerset1991,Skopal2005}) for the blackbody fit to the GALEX UV fluxes, we obtain a WD temperature of $2\times10^5$\,K and a luminosity of $1.4\times10^4$\,L$_\odot$. On the other hand, fitting blackbody models with radii of 0.2\,R$_\odot$ and 0.3\,R$_\odot$ (radii observed for significantly inflated WD atmospheres in symbiotics; e.g., Draco C1 in \citealt{Lewis2020} or LMC S63 in \citealt{Muerset96}) to the WD SED yields temperatures of 75,000\,K (1100\,L$_\odot$) and 47,500\,K (410\,L$_\odot$), respectively. The derived temperatures and luminosities for the assumed 0.1, 0.2, and 0.3\,R$_\odot$ models coincide with those values for other well-studied non-novae symbiotics (\citealt{Merc2019}, and references therein). Though we have no independent measure of the WD temperature (e.g., there are no X-ray data available for this system), we can be confident that the WD temperature is likely to be $<2\times10^5$\,K.

It should be noted that the temperature for the RG derived from the APOGEE spectra is slightly cooler than that from previous estimates, which put the giant at a temperature of $\sim$3850\,K \citep{Muerset96,Akras2019}. In fact, this previously derived temperature yields a better fit ($\chi^2 = 3.2$) to the SED than does the APOGEE temperature; however, because the APOGEE spectra provide significantly better spectral resolution than does a simple SED, the former likely provides the more trustworthy 
measure of the $T_\mathrm{eff}$. It should also be noted that the photospheric $T_\mathrm{eff}$ ($\sim$3850\,K) provided by modeling the SED depends mainly on how the SED of the given source of radiation is defined, and could provide an incorrect spectral type for the RG. The atmospheric $T_\mathrm{eff}$ provided by the APOGEE spectra depends on a selected spectral region contributed only by the RG, which avoids this issue. In the end, our derived temperature and radius for the WD are within the large range of values derived by prior work \citep{Muerset96,Morgan1992}.

\subsection{The Joker Results} \label{subsec:jokerres73}

\begin{figure*}[ht!]
\centering
\includegraphics[trim={0 0 2cm 1.1cm}, clip, height=0.30\textwidth]{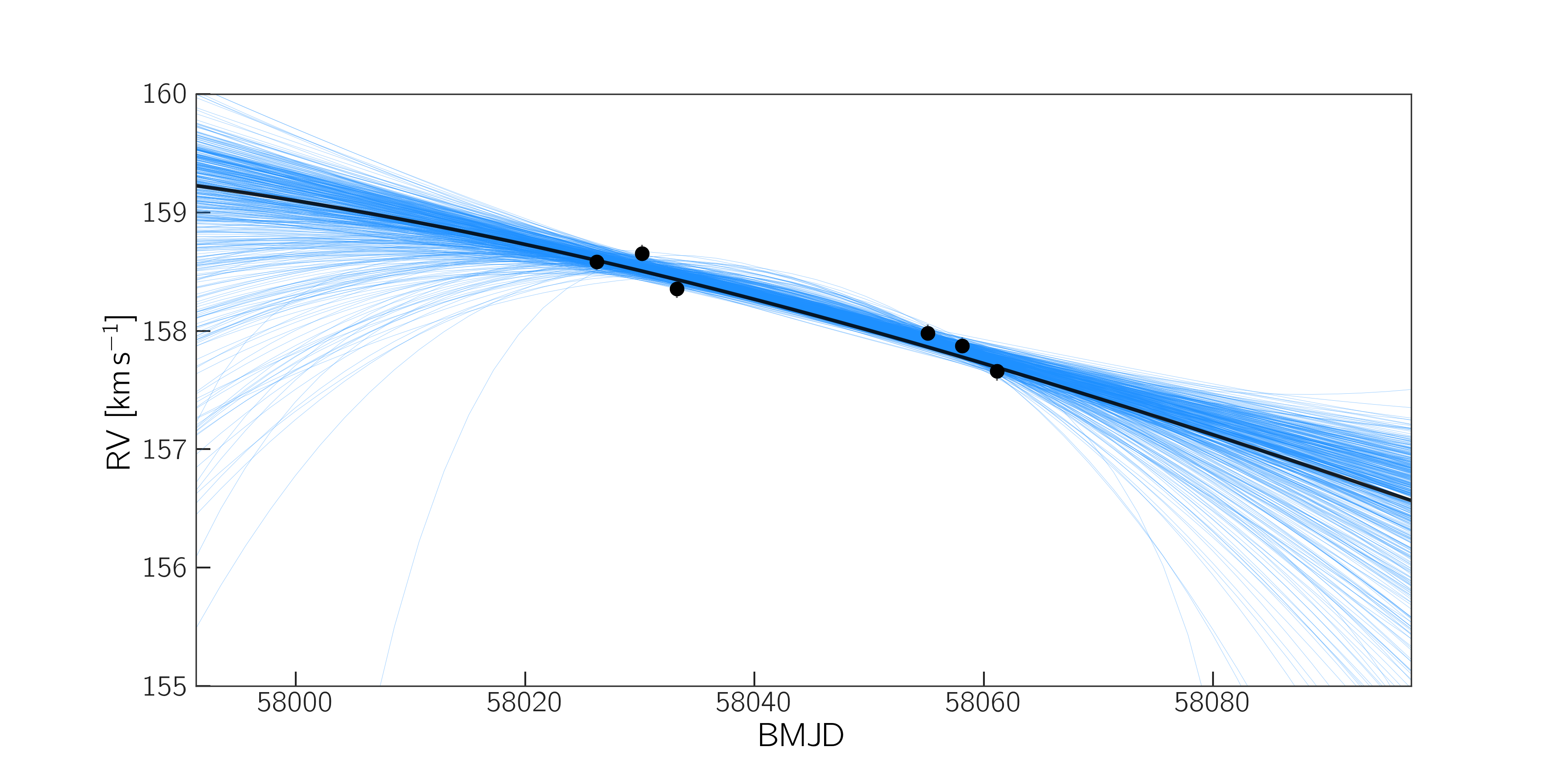}
\includegraphics[height=0.30\textwidth]{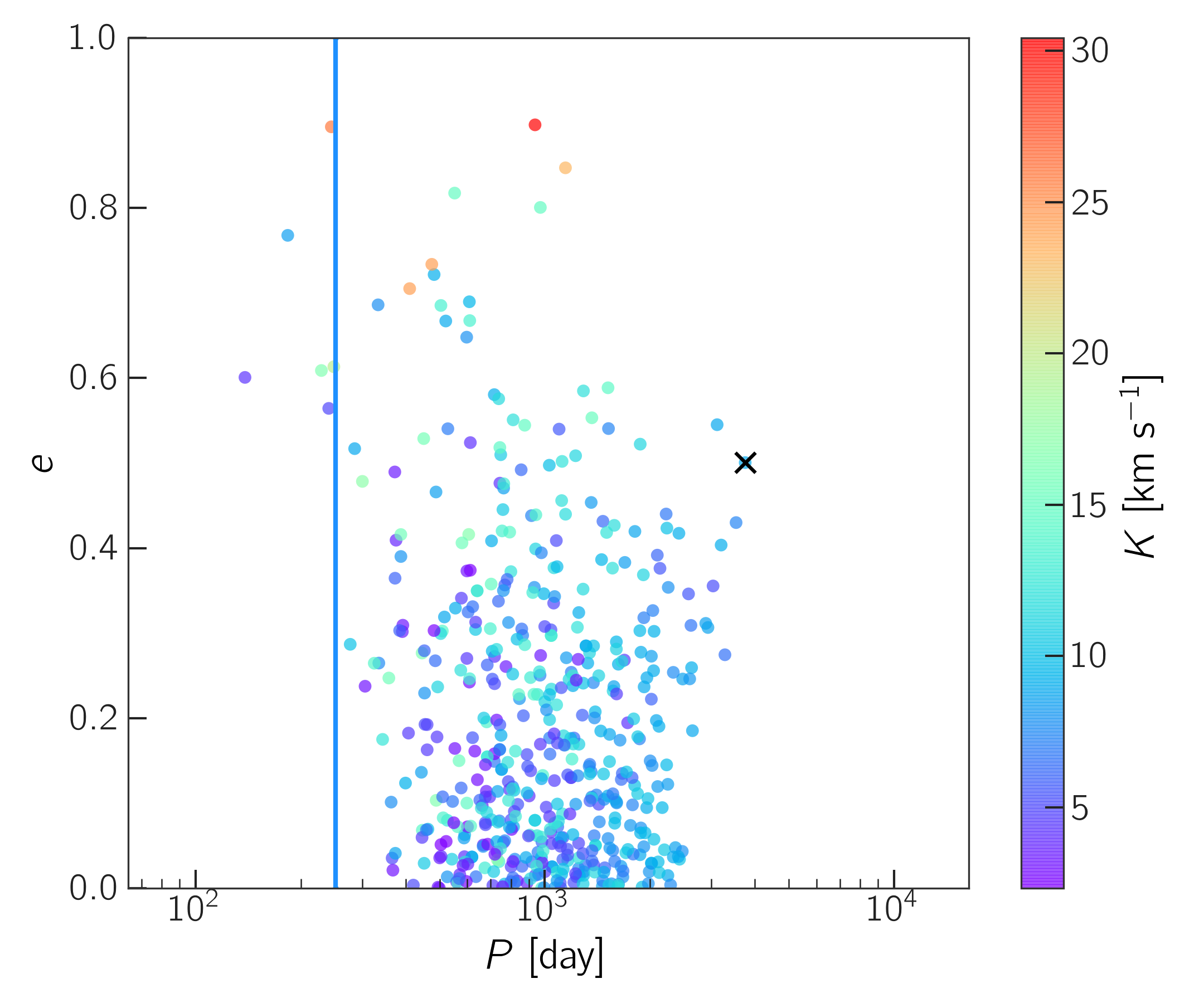}
\caption{\textit{The Joker} results for SMC N73. Left: APOGEE visit velocity data (black markers; error bars are shown, but are typically smaller than the marker) and the orbits computed from the posterior samples (blue lines). The MAP sample is shown by the black line. Right: Projections of the 512 posterior samples in period $P$ and eccentricity $e$, colored as a function of semi-amplitude $K$. Here, the MAP sample is marked by the black cross.}
\label{fig:jokers73}
\end{figure*}

\begin{figure*}[ht!]
\centering
\includegraphics[trim={1cm 1cm 2.5cm 3.5cm}, clip, width=0.65\textwidth]{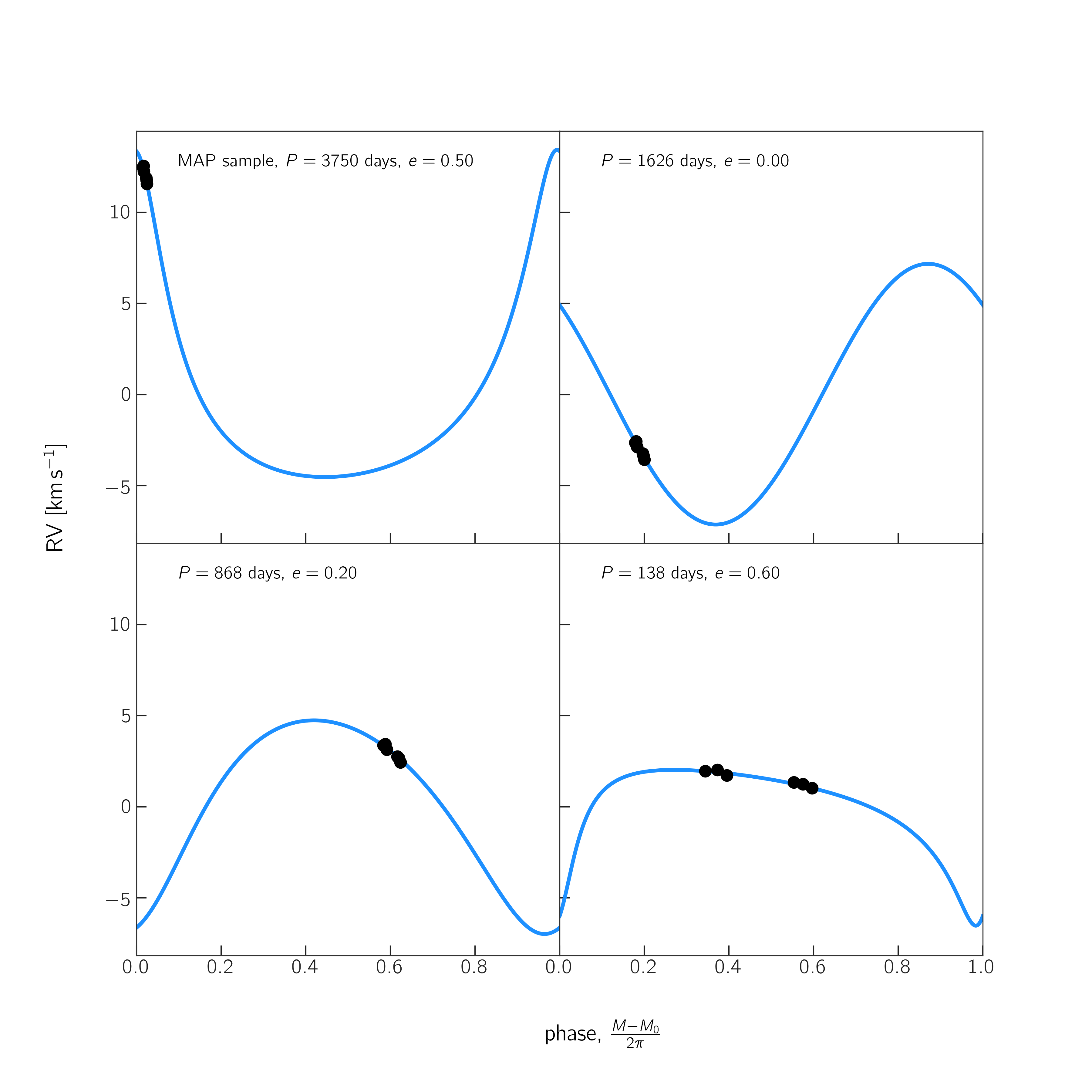}
\caption{Random selection of phased orbit solutions from {\it The Joker} for SMC N73 to show the variety of potential fits to the APOGEE visit velocity data (black markers). Orbits computed from the MAP sample ({\it upper left panel}) and three other possible period-eccentricity combinations are shown (blue lines).}
\label{fig:jokers73 phased}
\end{figure*}

\textit{The Joker} does not constrain the orbital period of the SMC N73 SySt well, as shown in Figure \ref{fig:jokers73},
likely because the orbital period of the system is much longer than the APOGEE observation baseline of only $\sim$30 days. It is unlikely that the orbital period is much shorter than the baseline of the APOGEE observations, as symbiotic systems tend to have orbital periods $>$200\,days \citep{mikolajewski2003}. In Figure \ref{fig:jokers73}, the maximum a posteriori (MAP) sample is indicated, though, given the sparse RV data for the the system, the MAP sample may not accurately represent the orbit of the system. For this reason, we show, in Figure \ref{fig:jokers73 phased}, classical phase diagrams for several other possible orbits spanning a range of orbital periods and eccentricities.

It is important to note that stellar jitter for a star with $\log{g} \sim 0$ may explain the $\sim$1\,km\,s$^{-1}$ RV variation observed for this star \citep{Hekker2008}. This is one reason why we are careful not to over-interpret the results of this orbital analysis.
 We only place lower limits on the period and semi-amplitude of the orbit, and give the median systemic velocity (the latter reported with 3$\sigma$ errors). Based on the results from \textit{The Joker}, we find $P > 270$\,days and $K > 2.5$\,km\,s$^{-1}$, where the lower limits on these parameters are defined by the value of the \nth{1} percentile of sample periods and semi-amplitudes, respectively. The derived systemic velocity, $v_0 = 154 \pm 13$\,km\,s$^{-1}$, is well within the velocity dispersion of stars in the SMC \citep{Harris2006}. 
It is not possible to place any constraints on the eccentricity of the orbit of SMC N73, because we do not have sufficiently constrained samples to prefer an eccentric over a circular orbit.

However, using the defined lower-limits for the period and semi-amplitude, and assuming $e = 0$, we can derive a lower-limit for the mass function $m_f$, where
\begin{equation}
m_f = \frac{M_\mathrm{WD}^3 \sin^3{i}}{\left( M_\mathrm{RG} + M_\mathrm{WD} \right)^2} =  \frac{PK^3}{2 \pi G} \left( 1 - e^2  \right)^{3/2},
\label{eq:mf}
\end{equation}
such that $m_f > 4.4\times10^{-4}$\,M$_\odot$.
Taking the limits of this function (i.e., for $M_\mathrm{RG} \gg M_\mathrm{WD}$ and $M_\mathrm{RG} \ll M_\mathrm{WD}$), it can be shown that for any inclination $i$, the absolute lower-limit on the mass of the WD is given as 
\begin{equation}
M_\mathrm{WD} > \max{\left( m_f, m_f^{1/3} M_\mathrm{RG}^{2/3} \right)}
\label{eq:min mass}
\end{equation}
\citep[e.g.,][]{podsiadlowski2014}. Substituting the lower-limit for $m_f$ and $M_\mathrm{RG} = 2.00$\,M$_\odot$, we find that the absolute lower-limit for the mass of the WD is $M_\mathrm{WD, min} > 0.17$\,M$_\odot$. However, this is not a very helpful lower limit, because this mass is equal to that of the least-massive known WD \citep[J0917+4638 with mass 0.17\,M$_\odot$, ][]{Kilic2007}, and, based on MIST stellar tracks, for a star with mass $\gtrsim$2\,M$_\odot$ on the main-sequence (MS) we expect the resulting WD to have a mass $\sim$0.6\,M$_\odot$ \citep{Dotter2016,Choi2016}. Because we do not present a full orbit for this system, and therefore the period and/or RV semi-amplitude may eventually be shown to be much larger with additional observations, the 
minimum mass derived in this work should serve only as absolute lower-limit on the mass of the SMC N73 WD.
The limits placed on the orbital parameters for SMC N73 are included in Table \ref{tab:params}.

\section{Analysis of LIN 358} \label{sec:lin358}

\subsection{Skopal (2015) SED}\label{subsec:sed358}

\citet{Skopal2015} provides an analysis of the SED for LIN 358 to determine the physical parameters of the white dwarf, giant, and nebular components of the system.
The method used is similar to that applied by \cite{Stassun2016}, but includes the nebular component $F_n$ in the net flux, such that
the total flux $F_\mathrm{tot}$ can be expressed as 
\begin{equation}
    F_\mathrm{tot}(\lambda)=F_\mathrm{h}(\lambda)+F_\mathrm{n}(\lambda)+F_\mathrm{g}(\lambda),
\end{equation}
where $F_\mathrm{h}$ is the component produced by the photosphere of the hot white dwarf and/or its disk, $F_\mathrm{n}$ is the nebular component produced by ionization of circumstellar material, and $F_\mathrm{g}$ is the flux produced by the giant. See \cite{Skopal2015} for additional details of the SED fit.

\citet{Skopal2015} found a giant temperature of $4000\pm200$ K, which is in good agreement with the APOGEE measurement of $3955\pm74$\,K and within the values derived by \citet{Muerset96} and \citet{Akras2019}. This agreement supports both the validity of the model presented by \citeauthor{Skopal2015} and the accuracy of the measurements made by APOGEE.
The radius of the giant derived by \cite{Skopal2015} is $178\pm18$\,R$_\odot$. Assuming the spectroscopic $\log{(g)} = 0.30 \pm 0.12$ derived from the APOGEE spectra (Section \ref{sec:data}), we then derive a mass of $2.31 \pm 0.79$\,M$_\odot$ for the giant.

For the WD, \citeauthor{Skopal2015} found the best fitting model to have a temperature of $(25 \pm 10) \times 10^4$\,K 
and radius $9.71 \pm 0.11$\,R$_\oplus$. 
This temperature is comparable to values derived by other studies, though the radius for the WD is marginally smaller than those obtained by the same analyses
\citep[e.g.,][]{Muerset96,Kahabka2006,Orio2007}.
This is likely because \citet{Skopal2015} considers a nebular contribution to the SED of LIN 358, whereas the previous literature did not,
so that the latter required a larger WD radius to account for the total flux. 
For the nebular contribution, \citet{Skopal2015} gives a temperature of $T_\mathrm{n} = (1.8 \pm 0.5) \times 10^4$\,K and an emission measure of $EM = (2.4\pm0.3) \times 10^{60}$\,cm$^{-3}$.

\subsection{The Joker Results} \label{subsec:jokerres358}

\begin{figure*}[ht!]
\centering
\includegraphics[width=0.49\textwidth]{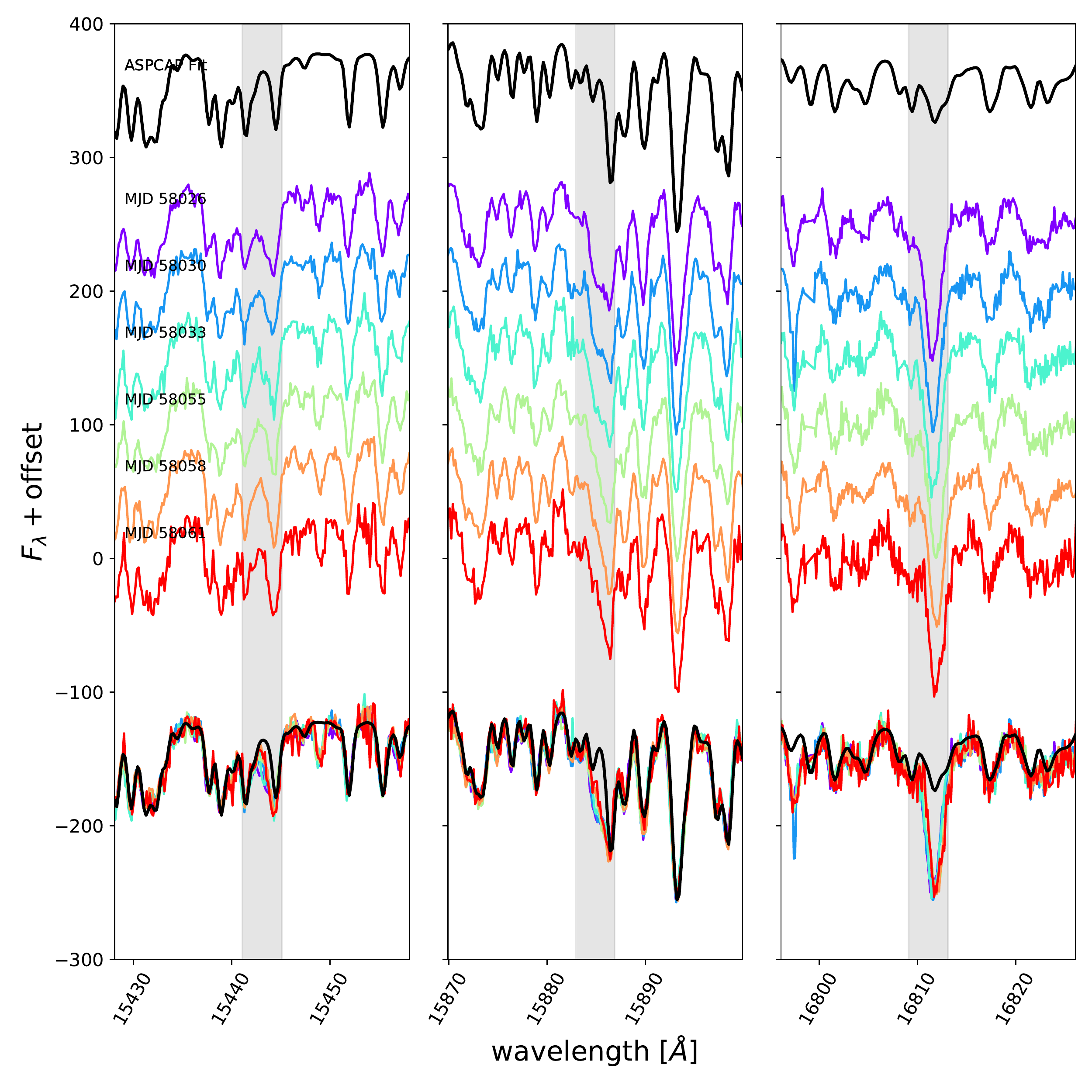} \hfill
\includegraphics[width=0.49\textwidth]{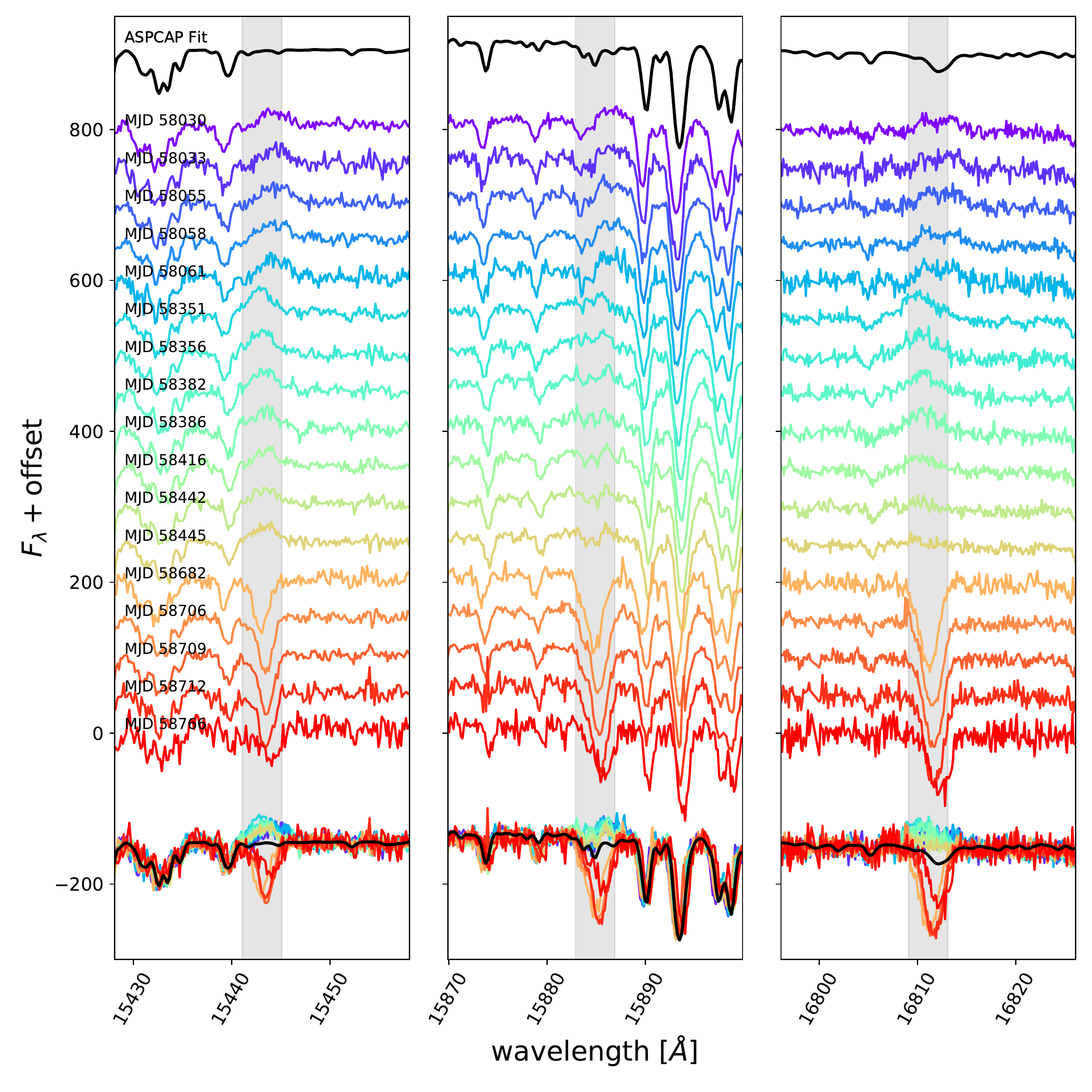}
\caption{Regions of the APOGEE spectra around three selected hydrogen Brackett series
lines (highlighted by the gray columns) from among the ten Brackett lines in the APOGEE spectra. Left: The SMC N73 visit spectra show deep Brackett absorption feature across all observations. Right: The same hydrogen features in the visit spectra of LIN 358 change from a broad emission feature at early epochs (MJD 58030 through 58445) to a narrower absorption feature in later observations (MJD 58682 and later). For both targets, the best-fiting ASPCAP spectrum is shown at the top (black) to highlight the fact that the metal lines in the visit spectra match the model, but that there is typically excess emission/absorption in the hydrogen lines for LIN 358, and in some cases for SMC N73.
The differences in the visit spectra from the ASPCAP results, as well as the variability in the Brackett lines, are highlighted by the superposition of all of the visit spectra onto the ASPCAP spectrum at the bottom of each panel.
The phenomena observed in the three Brackett lines shown for each system are representative of what is observed across all hydrogen lines in the APOGEE spectra of these two systems.}
\label{fig:spectra}
\end{figure*}

Following a visual inspection of metal lines in the APOGEE spectra, we discovered that
when shifted to rest-wavelength, the RVs derived by the APOGEE pipeline for five of the visits (MJD 58682 through 58766) shifted the 
spectra too far to the blue. 
We also found that in the spectra with the inaccurately derived RVs
the character of some of the spectral lines had changed.  In particular, prior to MJD 58682, several hydrogen Brackett 
lines were expressed as broad emission features, but in all subsequent spectra
they turn into 
narrow absorption lines (see Figure \ref{fig:spectra}). In Figure \ref{fig:spectra}, the best globally fitting ASPCAP spectrum is also shown (black) to highlight the fact that the metal lines in the visit spectra have the expected strengths (i.e., follow the model), but, in contrast, the hydrogen lines do not conform to expectation, with, in most cases for the two systems studied here, the Brackett lines showing emission or deeper absorption lines than expected. 
These differences, and, for LIN 358 strong variability in these differences, are highlighted at the bottom of Figure \ref{fig:spectra}, which shows the visit spectra superposed on the best global-fit,
ASPCAP model spectrum.

Because the varying spectral character is clearly affecting the derived APOGEE RVs for LIN 358, 
we rederived the visit-level RVs for this source using only eight metal lines in the APOGEE spectra (Fe I 15211.686, Fe I 15298.742, Fe I 15339.214, Fe I 15625.823, Fe I 15636.221, Mg I 15745.017, Mg I 15753.291, and Mg I 15770.150 \AA). The visit spectra shown in Figure \ref{fig:spectra} are shifted to the newly derived RVs, reported in Table \ref{tab:lin358RVs}, which are given by the mean of the eight individual RVs derived from each of the eight listed metal lines. We also show, in Appendix \ref{sec:appendix}, the portion of each visit spectra spanned by the metal lines used to derive the Rv of the RG component, to highlight the good agreement between the model and the observed spectra given the newly derived RVs.
In this figure, we highlight just three hydrogen Brackett lines (specifically, the 11--4, 14--4, and 17--4 transitions) even though a transition from broad emission to narrow absorption is observed for all hydrogen lines in the APOGEE spectra.

While the spectral variability observed in LIN 358 differs from the lack of such variability in the APOGEE spectra of SMC N73, 
the latter only span a temporal baseline of $\sim$30\,days (whereas the APOGEE observations of LIN 358 span a baseline of $\sim$2\,years), which is much shorter than the typical orbital period for symbiotic stars \citep{mikolajewski2003}. It is possible that longer term monitoring
of SMC N73 might reveal variability in the $H$-band hydrogen lines.
Presently we are not aware of any other giants observed by APOGEE that show 
such variability in the hydrogen lines like that seen in LIN 358. However, because spectral variability like this might be a clue to identifying previously unknown symbiotic stars in the survey, a thorough search for such systems is ongoing (Anguiano et al., in prep.).

\begin{figure*}[ht!]
\centering
\includegraphics[trim={0 0 2cm 1.1cm}, clip, height=0.30\textwidth]{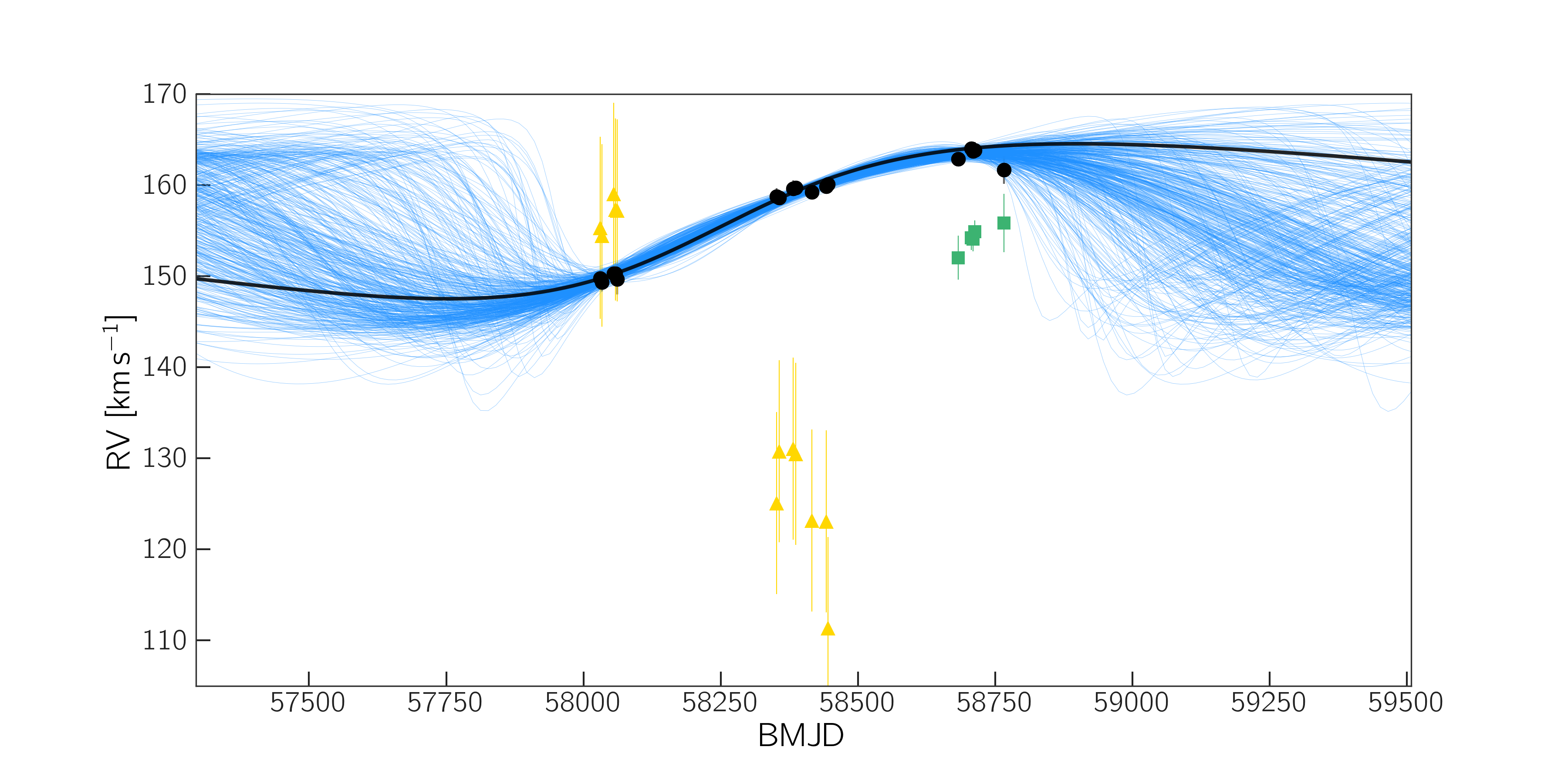}
\includegraphics[height=0.30\textwidth]{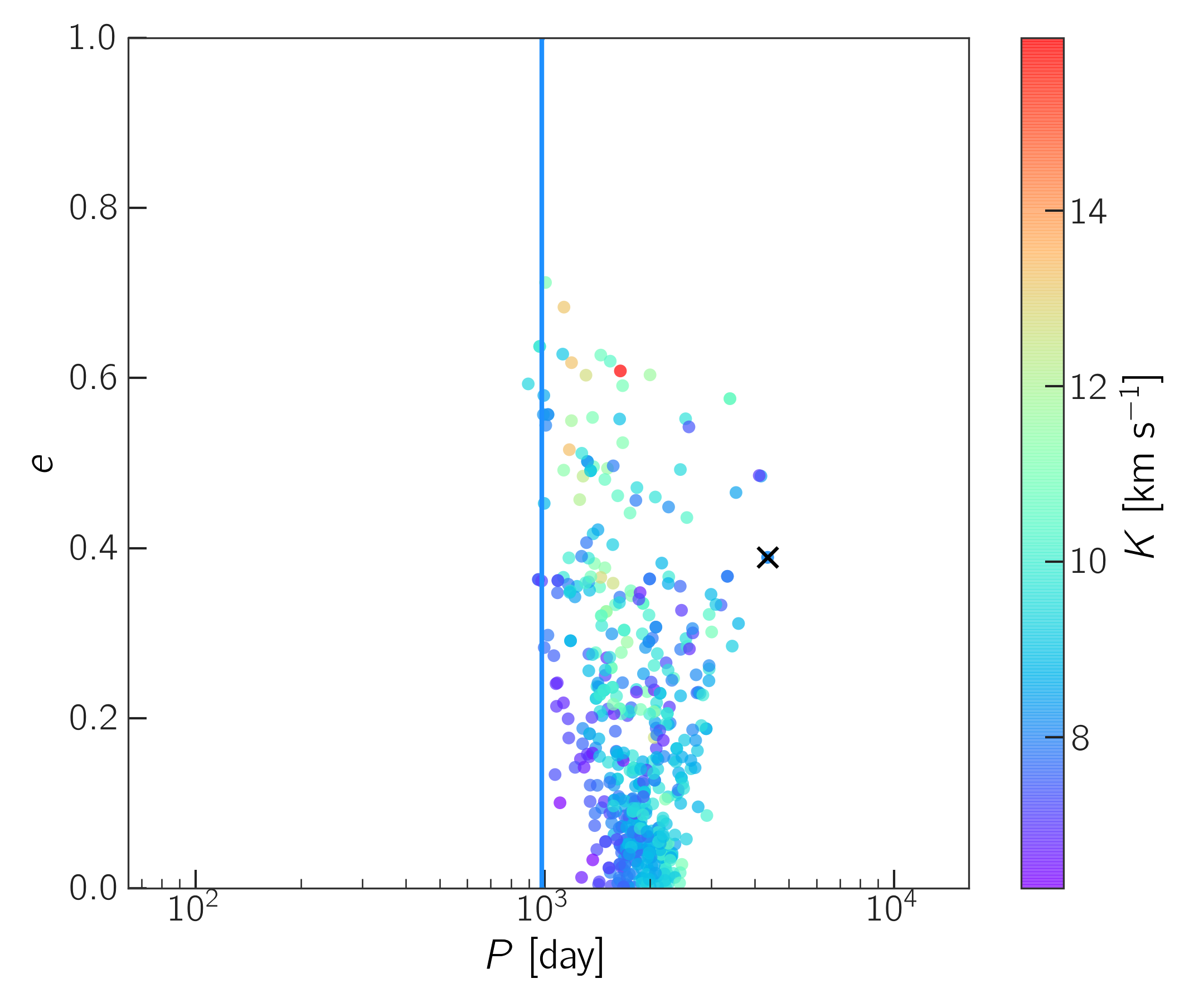}
\caption{\textit The RV variability and {\it The Joker} orbit-fitting results for LIN 358. Left: Visit velocity data derived from the metal lines in the APOGEE spectra (black markers; error bars are shown, but are typically smaller than the marker) and the orbits computed from the posterior samples (blue lines). The RVs derived from the Br11 line are also shown (yellow triangles if Br11 observed in emission, green squares if observed in absorption). The MAP sample is shown by the black line. Right: Projections of the 512 posterior samples in period, $P$, and eccentricity, $e$, given by {\it The Joker}. The MAP sample is marked by the black cross.}
\label{fig:jokers358}
\end{figure*}

RVs are derived from the Brackett lines by fitting either Gaussians to the double emission peaks observed in spectra taken prior to MJD 58682, taking the average as the line center used to calculate the RV,
or by fitting a single Gaussian to each of the ten hydrogen Brackett absorption lines observed in APOGEE spectra taken on/after MJD 58682, and taking the average of the measured RVs.  In either situation, RVs obtained from the Brackett lines differ significantly from the RVs derived from the metal lines of the giant star, as shown in Figure \ref{fig:jokers358}, where the RVs from the metal lines are shown as black dots and the RVs derived from the Br11 line are shown in color 
(yellow triangles for RVs derived when Br11 is in emission, green squares for RVs derived when it is in absorption). That the Brackett line RVs are systematically offset from the metal line RVs supports the notion that the former are 
related to the nebular component of the LIN 358 symbiotic system, which \citet{Skopal2015} has shown to contribute 
significantly to the SED of the system at infrared wavelengths (specifically, in the $JHK$-bands). Additionally, the hydrogen absorption feature likely originates (at least partially) from the outer, expanding atmosphere or wind of the RG component, and therefore can be blueshifted---typically by $\sim$ a few km\,s$^{-1}$---relative to the velocity of the RG \citep[e.g.,][]{shagatova2020}.

\begin{figure*}[ht!]
\centering
\includegraphics[trim={1cm 1cm 2.5cm 3.5cm}, clip, width=0.65\textwidth]{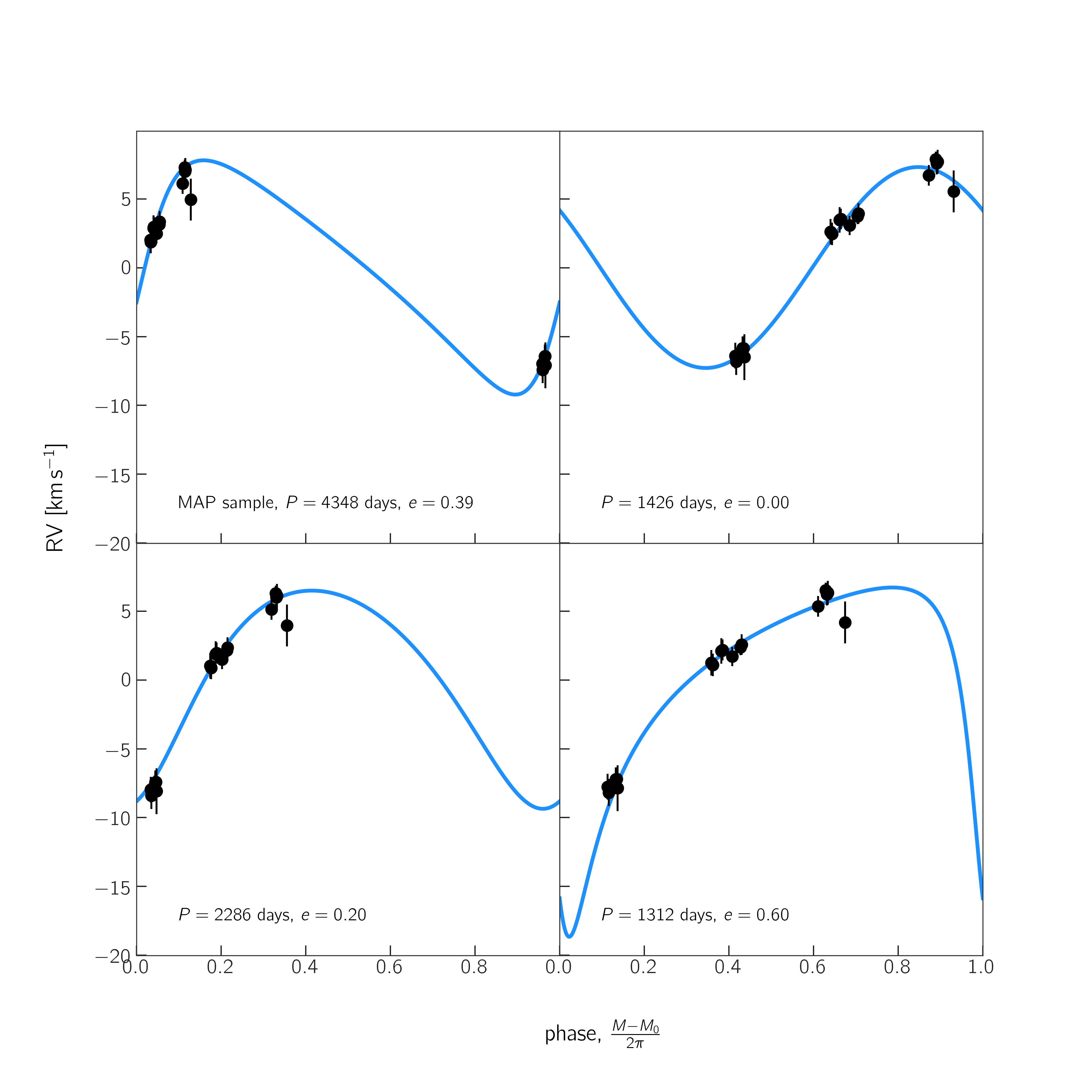}
\caption{A selection of phased orbit solutions from {\it The Joker} to the APOGEE visit velocity data (black markers) for LIN 358.  Orbits computed from the MAP sample ({\it upper left panel}, and three other possible period-eccentricity combinations are shown (blue lines).}
\label{fig:jokers358 phased}
\end{figure*}

Figure \ref{fig:jokers358} shows that the cadencing and long temporal baseline of the APOGEE RVs lead to solutions from \textit{The Joker} for LIN 358 that are better constrained than the solutions returned for SMC N73. Note, \textit{The Joker} is run on the RVs derived from the metal lines, not on the RVs derived from the Br11 absorption features. Again, in Figure \ref{fig:jokers358}, we show the MAP sample, as well as phase-folded RV curves for the MAP sample and several other possible orbital solutions in Figure \ref{fig:jokers358 phased}.

As with SMC N73, the returned sample RV curves do not completely constrain the orbit, so we only place lower limits on the period and semi-amplitude, and give the 99\% confidence interval for the systemic velocity. We find that the orbital period, $P$, must exceed $980$\,days and the RV semi-amplitude has $K > 6.5$\,km\,s$^{-1}$. The implied period for this system ($>$2.5\,years) is reasonable, given the typical range of 1 to 3 year periods for S-type symbiotics \citep{Skopal2005}. The kinematically derived period is also roughly in agreement with the period derived by \cite{Kuuttila2020}, who assumed maximal accretion efficiency via WRLOF, and found a period of $\sim$2.9\,yrs. The system barycenter velocity, derived from the sample RV curves returned by \textit{The Joker}, is equal to $156.2 \pm 5.3$\,km\,s$^{-1}$.
While we cannot use the APOGEE RVs to place limits on the eccentricity of the system, based on the spectral variability observed in the $H$-band hydrogen lines, we can assume that it is non-zero.
The limits placed on the orbital parameters for LIN 358 based on the APOGEE data are included in Table \ref{tab:params}.
From Equation \ref{eq:mf}, we find that the minimum mass function for the LIN 358 symbiotic binary is $m_f > 0.026$\,M$_\odot$, and from Equation \ref{eq:min mass}, that the absolute minimum mass of the WD (given any inclination angle $i$) is $M_\mathrm{WD,min} > 0.57$\,M$_\odot$.

\begin{table*}
\centering
\caption{Parameters of SMC N73 and LIN 358. \label{tab:params}}
\begin{tabular}{lcccl}
\hline
\hline
Parameter & SMC N73 & LIN 358 & Units & Reference \\
& (2M01043930-7548247) & (2M00591224-7505176) & &\\
\hline
\multicolumn{5}{l}{Red Giant Parameters} \\
$T_\mathrm{eff}$ & $3590\pm50$ & $4000\pm200$ & K & APOGEE DR16, \citet{Skopal2015} \\
$\log{(g)}$ & $0.10\pm0.10$ & $0.30\pm0.12$ & cgs & This work \\
$\mathrm{\left[Fe/H\right]}$ & $-0.75\pm0.12$ & $-1.00\pm0.10$ & dex & This work \\
$R_\mathrm{RG}$ & $208 \pm 16$ & ${178}\pm{18}$ & R$_\odot$ & This work, \citet{Skopal2015} \\
$M_\mathrm{RG}$ & ${2.00}\pm{0.55}$ & ${2.31}\pm{0.79}$ & M$_\odot$ & This work \\

\multicolumn{5}{l}{White Dwarf Parameters} \\
$T_\mathrm{eff}$ & $<2\times10^5$ & $\left(2.5\pm0.1\right)\times10^5$ & K & This work, \citet{Skopal2015} \\
$R_\mathrm{WD}$ & $\sim 10.9$ & $9\pm2$ &  R$_\oplus$ & This work, \citet{Skopal2015} \\
$M_\mathrm{WD, min}$ & $>0.17$ & $> 0.57$ & M$_\odot$ & This work \\

\multicolumn{5}{l}{Nebula Parameters} \\
$T_\mathrm{eff}$ & --- & $\left(18\pm5\right)\times10^3$ & K & \citet{Skopal2015}\\
EM & --- & $2.4\pm0.3$ & $10^{60}$ cm$^{-3}$ & \citet{Skopal2015}\\

\multicolumn{5}{l}{System Parameters} \\
$d_\mathrm{SMC}$ & \multicolumn{2}{c}{${63.4}\pm{2.3}$} & kpc & \cite{distmod} \\
$P$ & $> 270$ & $> 980$ & days & This work \\
$e$ & --- & $> 0$ & & This work  \\
$K$ & $> 2.5$ & $> 6.5$ & km\,s$^{-1}$ & This work \\
$v_0$ & $154 \pm 13$ & $156.2 \pm 5.3$ & km\,s$^{-1}$ & This work \\
\hline
\hline
\end{tabular}
\end{table*}

\section{Discussion}
\label{sec:discussion}
Both symbiotic systems explored here show some unusual characteristics.  For example, whereas most SySts exhibit nebular emission lines, SMC N73 shows distinctly stronger than expected absorption for some Brackett lines.  Meanwhile, LIN 358 stands out even more by exhibiting highly variable hydrogen features that, over the course of two years, were observed to evolve from broad, but weak, emission to stronger than expected absorption.
While it is not unusual for symbiotic stars to show variable emission line strengths, particularly in the hydrogen series (as observed for, e.g., HD 4174 by \citealt{Smith1980}, AG Dra by \citealt{Leed2004}, Nova Sco 2015 by \citealt{Srivastava2015}, and SU Lyn by \citealt{ Mukai2016}), only a handful of systems are known to show hydrogen in {\it both} emission and absorption over the course of their orbits \citep[e.g.,][]{Bensammar1988,Bensammar1989,munari1993,Schmutz1994,Schild1996,shagatova2020}.  Though the nature of the strong hydrogen (specifically, H$\alpha$) absorption observed for these systems---many of which are eclipsing systems---is not well understood, absorption in this line is always observed at or near the inferior conjunction of the giant \citep[e.g.,][]{shagatova2020}. This fact is likely the result of varying optical depths at different orbital phases for a given hydrogen transition. As shown by \cite{Seaquist1984}, during quiescent phases, the environment surrounding the symbiotic is divided into an ionized hydrogen region, around the WD, and a neutral region, surrounding the giant; in this case, at phases in the orbit where the WD is between the RG and the observer, strong hydrogen emission is observed in the spectra, and when the RG is closer to the observer, stronger absorption features are observed. Though we currently have no evidence that LIN 358 is an eclipsing symbiotic, given the available data, this is one interesting possible explanation for the observed line variability.

The change in strength of these nebular features might also indicate variability in the mass accretion of the WD, which, in turn, may be a reflection of a non-circular binary orbit, since the instantaneous mass-accretion rate would be expected to vary strongly with the orbital velocity and separation of the stars. For example, the hydrodynamical simulations of AGB binaries by e.g., \citet{Saladino2019} have shown that for high eccentricity systems one should expect 
significant variability in the mass accretion rate of the WD, and therefore significant variability in the nebular emission features, as observed here in the Brackett lines and found for other lines in LIN 358 spectra, like H$\alpha$ (see discussion in Section \ref{sec:introLIN358}). A highly eccentric orbit could also cause variable X-ray emission as the WD accretes at different rates, and a long-term change in X-ray flux has been reported for LIN 358 by \citet{Kahabka2006}.
Constraining both the orbital period and inclination, and determining whether the orbital phases map directly onto regular variations in nebular line shape and strength, 
will be important steps for understanding more fully the architecture of this interesting symbiotic system. 

Because LIN 358 is a known SSS, which requires steady nuclear burning of accreted hydrogen on the surface of the WD, if a bolometric luminosity of the WD is assumed, the accretion rate can be constrained. For example, \cite{Kuuttila2020} find a high WD luminosity of $1.02 \times 10^{38}$\,erg\,s$^{-1}$, which requires a similarly high accretion rate of $6\times10^{-7}$\,M$_\odot$\,yr$^{-1}$. However, given the derived mass-loss rate of the giant ($\sim10^{-6}$\,M$_\odot$\,yr$^{-1}$; \citealt{Kuuttila2020}), such a high accretion rate cannot readily be explained by standard Bondi-Hoyle-Lyttleton wind accretion \citep[which has an efficiency of only a few percent;][]{BHL1939,BHL1944,BHL1952}, suggesting that a mass transfer mechanism that is more efficient than standard Bondi-Hoyle wind accretion is required.
This is consistent with other studies that find giant stellar radii that are too small for RLOF as the typical mass transfer mechanism in SySts \citep{Mikolajewska2007,Munari2019}. For example, \cite{Lewis2020} found that the radius of the RL is a factor of $\sim$2$\times$ too large for standard RLOF to occur in the Draco C1 symbiotic system, so that the atmosphere of the giant is contained within the RL, and therefore mass transfer must occur via some other mechanism. However, the possibility of RLOF cannot be ruled out entirely; \citet{boffin2014}, for example, find that, in a sample of six mass-transferring RGs, 50\% have RL filling factors close to unity. Given the relatively small sample sizes presented in these studies, a larger sample of symbiotic systems with well-characterized geometries is necessary to rule out RLOF as an alternate mass transfer mechanism.

\begin{figure}[h!]
\centering
\includegraphics[width=\columnwidth]{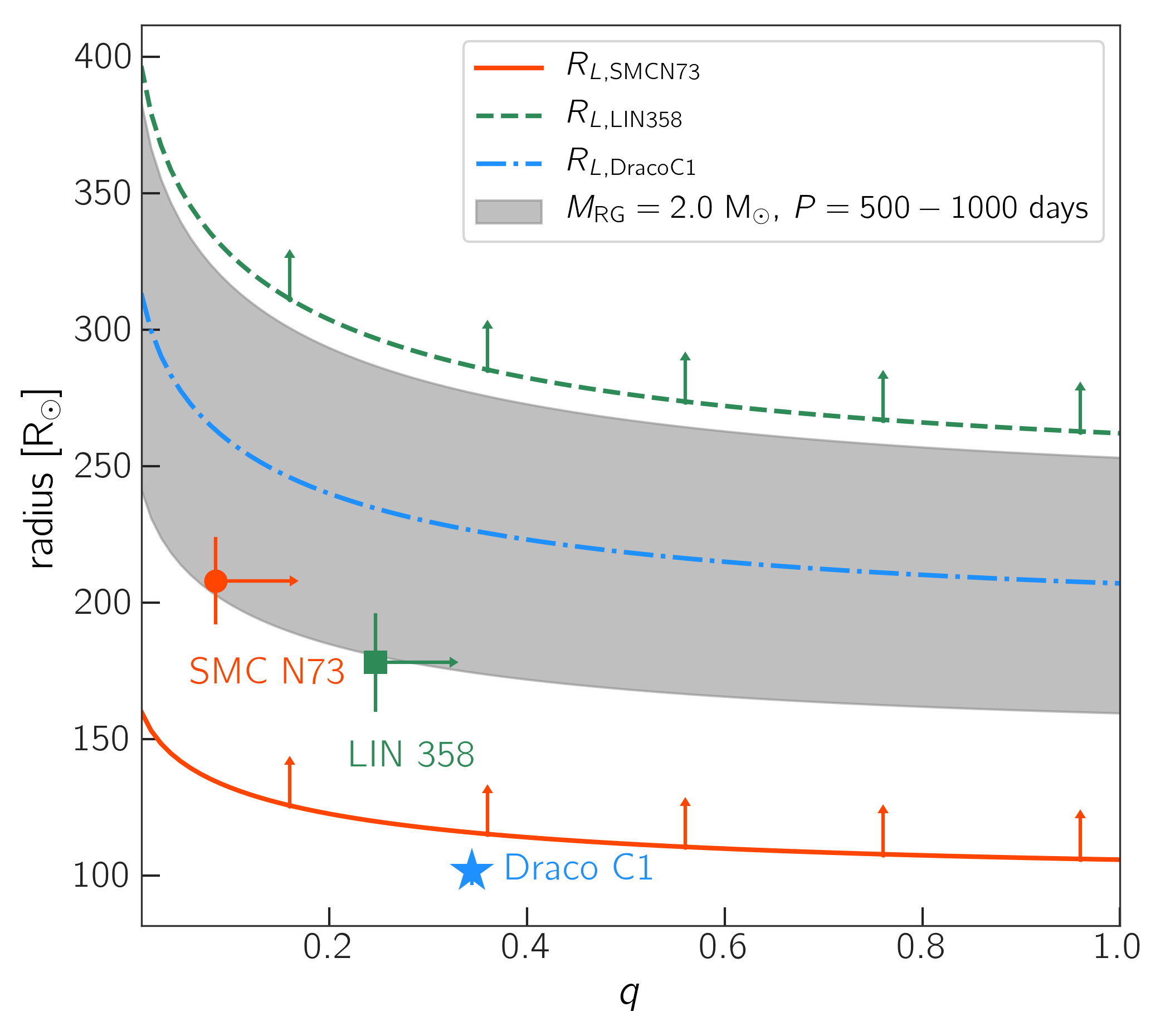}
\caption{Radius of the Roche lobe, $R_L$, of the RG components of the symbiotic systems, compared to the radius of the RG photosphere, $R_\mathrm{RG}$. The RL radii (indicated by the red solid and green dashed lines) are based on the lower-limits placed on the orbital periods for these two systems, and are shown as a function of the mass ratio, $q = M_\mathrm{WD}/M_\mathrm{RG}$ (Equation \ref{eq:RL}). The radii of the photosphere of SMC N73 and LIN 358 versus their minimum mass ratios, $q = 0.060$ and $0.225$, are indicated by the red circle and green square, respectively. We also show RL radii for a 2.0\,M$_\odot$ primary with orbital periods ranging from 500\,days (lower-edge of the grey area) to 1000\,days (upper edge of the grey area), to show the range of possible RL radii if SMC N73 is found to have a period much longer than 270\,days. For reference, the same parameters are plotted for Draco C1 (blue dot-dashed line and blue star).}
\label{fig:RL}
\end{figure}

Following the formulation laid out by \cite{Lewis2020}, we calculate the RL radius for the RG components, $R_L$, of both systems as a function of RG mass ($M_\mathrm{RG}$),  mass ratio ($q = M_\mathrm{WD}/M_\mathrm{RG}$), and orbital period ($P$), such that
\begin{equation}
    R_L = \frac{0.49q^{-\frac{2}{3}}}{0.6q^{-\frac{2}{3}} + \ln{(1 + q^{-\frac{1}{3}})}} \times \left( \frac{G M_\mathrm{RG} P^2 (1+q)}{4 \pi^2} \right)^{\frac{1}{3}},
    \label{eq:RL}
\end{equation}
where the right-most fraction represents the minimum orbital separation $a$, 
assuming a circular orbit \citep{Eggleton1983}.
In Figure \ref{fig:RL}, based on the RG masses and lower-limits placed on the orbital periods for the systems in this work, we apply Equation \ref{eq:RL} to show the RL radius for SMC N73 and LIN 358 (given by the solid red and green dashed lines, respectively) for a range of mass ratios, $q = 0.01$ to $1.0$. Note, because we have not observed a complete orbit for either of these systems, and the RL radius is most strongly dependent on the orbital period (i.e., $R_L \propto P^{\frac{2}{3}}$), the lines represent lower-limits for the RL radius (as indicated by the vertical arrows). In the figure, the red circle and green square indicate the derived $R_\mathrm{RG}$ and lower-limit placed on the mass ratio $q = M_\mathrm{WD, min}/M_\mathrm{RG}$ in this work for the SMC N73 and LIN 358 symbiotics, respectively.

Based on this information, we conclude that the LIN 358 symbiotic binary must have an orbital separation $a$ (and therefore $R_L$) that is too large for the standard RLOF scenario (i.e., $R_\mathrm{RG} < R_L$ for LIN 358). Instead, hydrodynamical simulations by e.g., \cite{mohamed2007,mohamed2012} show that such systems may undergo wind Roche lobe overflow (WRLOF), where the star itself does not fill the RL but the stellar wind does.
In this scenario, the stellar wind is focused towards the binary orbital plane, allowing for more efficient mass transfer than can be explained by standard RLOF \citep{devalborro2009}. 

While we conclude that the LIN 358 symbiotic is likely undergoing mass transfer via WRLOF, the accretion mechanism for SMC N73 is still very uncertain. 
Based on the lower-limit placed on the orbital period derived from the APOGEE RVs---which span a temporal baseline of only $\sim$30\,days---it appears that the radius of the giant component of this symbiotic exceeds its RL radius, and therefore is accreting via standard RLOF. However, it is important to again note that the APOGEE RVs do not show any inflection in their trend
and so are likely representative of (at most) $\sim$half the true orbital period of the system. For this reason, we show possible RL radii for a 2.0\,M$_\odot$ primary with orbital periods ranging from 500\,days ($\sim$2$\times$ period derived in this work, lower-edge of the grey area) to 1000\,days ($\sim$4$\times$ period derived in this work, upper edge of the grey area) in Figure \ref{fig:RL}. If the true orbital period of SMC N73 exceeds $\sim$500\,days, the RL radius will be greater than $R_\mathrm{RG}$, and we would assume that the system is accreting via WRLOF. Given the relatively short baseline of the APOGEE observations for SMC N73 (compared to the typical period of symbiotic binaries), we do not have sufficient evidence at this time to constrain the mode of mass transfer for this system.

However, a more interesting lower-limit on the period of this system can be derived if we assume that RLOF is the mass transfer mechanism for the SMC N73 symbiotic (i.e., $R_\mathrm{RG} = R_L$). Assuming $M_\mathrm{RG} = 2.0\,M_\odot$, $M_\mathrm{WD} = 0.6\,M_\odot$, and $R_\mathrm{RG} = 208\,R_\odot$, we can derive the maximum orbital period for which RLOF to occur. By substituting for $q = 0.3$ and $M_\mathrm{RG}$ in Equation \ref{eq:RL}, and setting the result equal to $R_\mathrm{RG}$, we find $R_L = 2.8\,R_\odot\,\mathrm{days}^{-2/3} \times P^{2/3} = R_\mathrm{RG}$. Solving for the orbital period, $P$, we find that for RLOF to occur the orbital period must equal $P \sim 640$\,days; for longer periods, $P \gtrsim 640$\,days, the RG will be contained within its Roche lobe, and some other mass transfer mechanism must be at play (i.e., WRLOF). This period agrees well with the known period distribution for symbiotic binaries \citep{mikolajewski2003}. However, this calculation requires us to rely strongly on the assumed masses of the components of this symbiotic. In particular, the WD mass assumed (0.6\,M$_\odot$) relies on the assumption that on the MS this star had a mass of 2.0\,M$_\odot$; while this assumption is approximately correct, the derivation of the WD mass via this method is not exact. For this reason, we assume the lower-limit for the period and mass ratio derived by the orbital analysis shown in Figure \ref{fig:jokers73}.

WRLOF requires a slowly accelerating wind, with acceleration radius $R_d \sim R_L$ \citep{abate2013}---a condition typically met for binaries containing Mira-type, AGB primaries, which have relatively small wind terminal velocities ($<$10\,km\,s$^{-1}$; e.g., \citealt{Wood2000,Matthews2006,Lee2007}). For S-type systems that contain normal giants,
the wind terminal velocity can be a few times larger, so the requirement for a slowly accelerating wind is not met. However, for such systems, 
an effective wind mass transfer can be caused by rotation of the mass-losing giant that leads to compression of the wind in the orbital plane \citep{SkopalCarikova2014,Shagatova2016}. Both SMC N73 and LIN 358 are S-type systems, but their giants are not quite normal, and we know very little about their winds and spectroscopic orbits, so both possibilities for enhanced mass transfer are viable.

\section{Summary}
\label{sec:summary}

By fitting the SEDs of these systems and applying APOGEE surface gravities obtained from
measurements of Ti I and Ti II lines to derived masses, we provide stellar parameters for the SMC N73 and LIN 358 symbiotic binaries.
SMC N73 is composed of a $\sim$3600\,K giant with radius $208 \pm 16$\,R$_\odot$ and mass $2.00 \pm 0.55$\,M$_\odot$, and the WD has a temperature of $(16.6 \pm 1.2) \times 10^4$\,K with radius $13 \pm 2$\,R$_\oplus$. While \citet{Skopal2015} previously derived the temperature and radius for both components of the LIN 358 binary, we also provide, from the APOGEE $\log{(g)}$, an estimate of the stellar mass of the RG component, $2.31 \pm 0.79$\,M$_\odot$. Because we have not observed a complete orbit for either of these systems, the orbital parameters (and therefore companion masses) derived by \textit{The Joker} are not well constrained and serve only as lower limits. For SMC N73, we find a period $>$270\,days and a minimum WD mass of 0.17\,M$_\odot$; for LIN 358, we find a period $>$980\,days and a WD mass of 0.57\,M$_\odot$. These values (for temperature, radius, mass, and period) generally agree with prior literature values for the giant and WD components of both systems.

Additionally, based on variability observed in the hydrogen Brackett lines of the APOGEE visit spectra, we claim that the LIN 358 system may have an eccentric orbit; however, without additional RV observations spanning a full orbital period, we cannot place further constraints on the eccentricity or the cause of the spectral variability. Similar variability is not observed for SMC N73, though there is additional absorption than what is to be expected for non-symbiotic, giant stars of the same temperature. We also conclude that LIN 358 is most likely transferring matter to the WD via WRLOF, though the mass transfer mechanism for SMC N73 is not clearly revealed by the available data.

Further observations of both systems, both photometric and spectroscopic, would be immensely beneficial for the complete classification of the stellar components that make up these symbiotic binaries. In particular, X-ray observations of SMC N73 would provide better constraints on the temperature and radius of the WD component, and spectroscopic follow-up of both systems over an entire orbital period---likely longer than 1000\,days---would allow precise derivation of orbital parameters, and provide additional insights to the wind accretion mechanisms occurring in these systems.

\acknowledgments

We would like to thank the helpful referee for their detailed comments and suggested additions; they significantly improved the presentation of this work.
J.E.W., H.M.L., B.A., and S.R.M. acknowledge support from National Science Foundation grants AST-1616636 and AST-1909497. B.A. also acknowledges support from an AAS Chrétien International Research Grant. C.A.P is thankful for funding from the Spanish government (AYA2017-86389-P). DAGH acknowledges support from the State Research Agency (AEI) of the Spanish Ministry of Science, Innovation and Universities (MCIU) and the European Regional Development Fund (FEDER) under grant AYA2017-88254-P.
This research made use of the New Online Database of Symbiotic Variables \citep{Merc2019} and is based on observations made with the Galex Evolution Explorer, obtained from the MAST data archive at the Space Telescope Science Institute, which is operated by the Association of Universities for Research in Astronomy, Inc., under NASA contract NAS 5–26555. 

Funding for the Sloan Digital Sky Survey IV has been provided by the Alfred P. Sloan Foundation, the U.S. Department of Energy Office of Science, and the Participating Institutions. SDSS-IV acknowledges
support and resources from the Center for High-Performance Computing at
the University of Utah. The SDSS web site is www.sdss.org.

SDSS-IV is managed by the Astrophysical Research Consortium for the Participating Institutions of the SDSS Collaboration including the Brazilian Participation Group, the Carnegie Institution for Science, Carnegie Mellon University, the Chilean Participation Group, the French Participation Group, Harvard-Smithsonian Center for Astrophysics, Instituto de Astrof\'isica de Canarias, The Johns Hopkins University, Kavli Institute for the Physics and Mathematics of the Universe (IPMU) / University of Tokyo, the Korean Participation Group, Lawrence Berkeley National Laboratory, Leibniz Institut f\"ur Astrophysik Potsdam (AIP), Max-Planck-Institut f\"ur Astronomie (MPIA Heidelberg), Max-Planck-Institut f\"ur Astrophysik (MPA Garching), Max-Planck-Institut f\"ur Extraterrestrische Physik (MPE), National Astronomical Observatories of China, New Mexico State University, New York University, University of Notre Dame, Observat\'ario Nacional / MCTI, The Ohio State University, Pennsylvania State University, Shanghai Astronomical Observatory, United Kingdom Participation Group, Universidad Nacional Aut\'onoma de M\'exico, University of Arizona, University of Colorado Boulder, University of Oxford, University of Portsmouth, University of Utah, University of Virginia, University of Washington, University of Wisconsin, Vanderbilt University, and Yale University.

This work has made use of data from the European Space Agency (ESA) mission {\it Gaia} (\url{https://www.cosmos.esa.int/gaia}), processed by the {\it Gaia} Data Processing and Analysis Consortium (DPAC, \url{https://www.cosmos.esa.int/web/gaia/dpac/consortium}). Funding for the DPAC has been provided by national institutions, in particular the institutions participating in the {\it Gaia} Multilateral Agreement. This research also made use of the SIMBAD database, operated at CDS, Strasbourg, France.

\vspace{5mm}

\software{astropy \citep{astropy},  
          thejoker \citep{Joker2017,jokercode2019,JokerVAC}, 
          }

\appendix
\setcounter{table}{0}
\renewcommand{\thetable}{A\arabic{table}}

\section{APOGEE and Derived RVs} \label{sec:appendix}

APOGEE RV errors are known to be underestimated \citep[e.g.,][]{Badenes18,Cottaar2014}; for this reason,  we apply the expression presented in Brown et al.~(in prep.) 
\begin{equation}
\sigma_\mathrm{RV}^2 = (3.5 \sigma^{1.2})^2 + (0.072\,\mathrm{km\,s}^{-1})^2,
\end{equation}
where $\sigma$ is the visit-level RV error and $\sigma_\mathrm{RV}$ is the total, inflated visit velocity error for a given visit. This essentially applies a $72$\,m\,s$^{-1}$ lower-limit for the visit-level RV uncertainties. Note, this expression is only applied to the visit uncertainties for SMC N73, as the RVs for LIN 358 were re-derived from the APOGEE visit spectra using other methods and therefore, do not have underestimated errors.

\begin{figure*}[ht!]
\centering
\includegraphics[trim={3cm 0 4cm 2cm}, clip, width=\columnwidth]{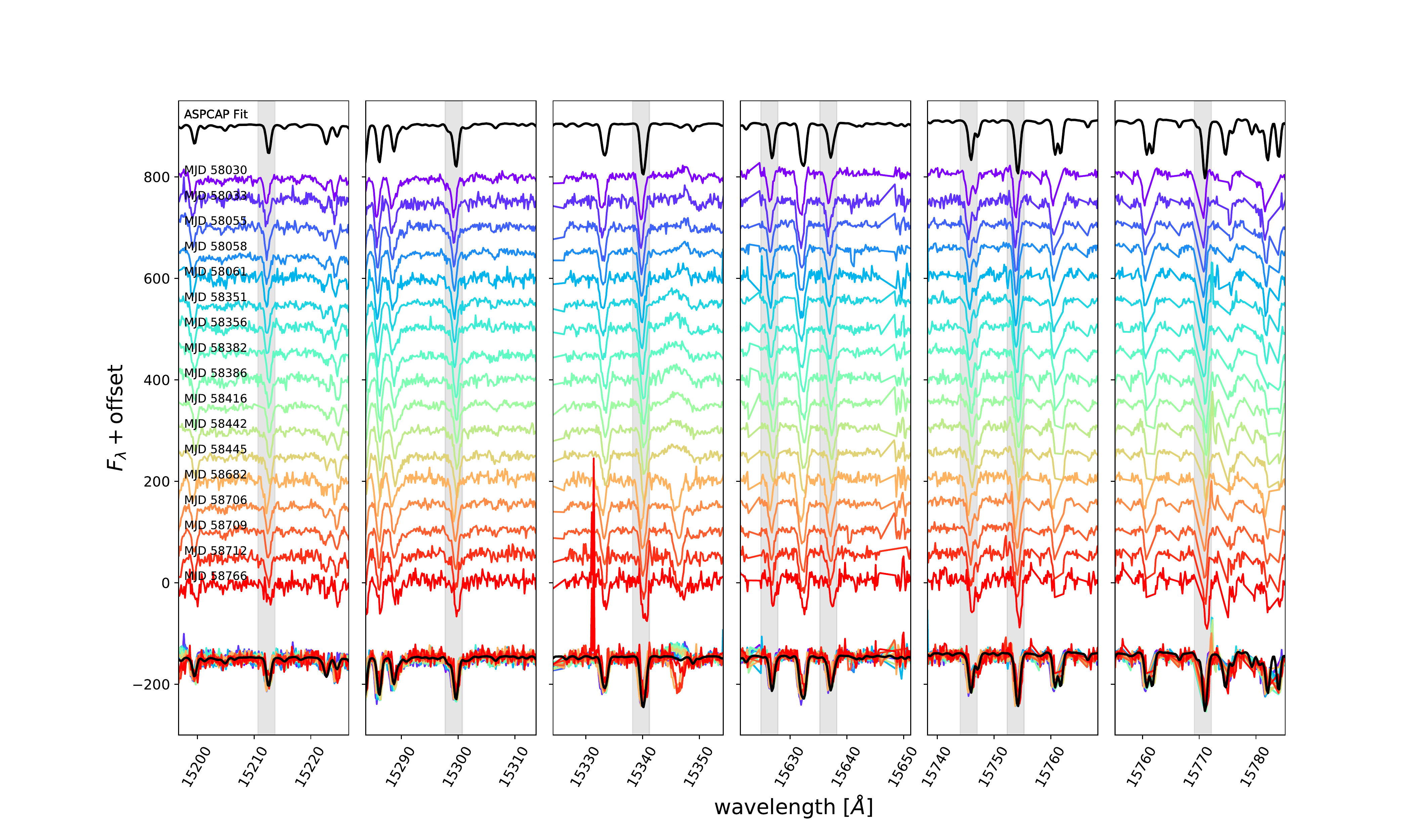}
\caption{Regions of the APOGEE visit spectra around the eight metal lines (highlighted by the gray columns) used to derive the RVs of the RG component of LIN358. The best-fitting ASPCAP spectrum is shown at the top (black), and again in the superposition of all spectra (bottom), to demonstrate the general goodness of fit of the model spectrum to the data.
Bad pixels, as defined by the APOGEE pixel-level bitmask, are masked in each visit spectra.}
\label{fig:metal lines}
\end{figure*}

The APOGEE pipeline-derived RVs for SMC N73 and the associated errors $\sigma_\mathrm{RV}$, are reported in Table \ref{tab:n73RVs}. The RVs and errors derived from the eight metal lines (listed in Section \ref{sec:lin358}) in the APOGEE spectra of LIN 358 are given in Table \ref{tab:lin358RVs}. In this case, the derived RVs are the mean of the RVs derived from fits to each of the eight metal lines. We also show, in Figure \ref{fig:metal lines}, the metal lines used to derived these RVs, to highlight the good agreement between the model and the observed spectra.

\begin{table*}[h]
\centering
\caption{RV measurements and associated errors from APOGEE for SMC N73. \label{tab:n73RVs}}
\begin{tabular}{ccc}
\hline
\hline
MJD & RV & $\sigma_\mathrm{RV}$\\
 & (km s$^{-1}$) & (km s$^{-1}$) \\
\hline

58026.22591 & 158.585 & 0.074  \\
58030.18268 & 158.653 & 0.075  \\
58033.23493 & 158.355 & 0.076  \\
58055.11245 & 157.980 & 0.076  \\
58058.09527 & 157.872 & 0.073 \\
58061.15527 & 157.661 & 0.082 \\
\hline
\end{tabular}
\tablecomments{This table is available in its entirety in machine-readable form.}
\end{table*}

\begin{table*}[h]
\centering
\caption{RVs and errors measured from metal lines and Br11 lines for LIN 358. \label{tab:lin358RVs}}
\begin{tabular}{ccccc}
\hline
\hline
MJD & RV & $\sigma_\mathrm{RV}$ & RV$_\mathrm{Br11}$ & $\sigma_\mathrm{RV, Br11}$\\
 & (km s$^{-1}$) & (km s$^{-1}$) & (km s$^{-1}$) & (km s$^{-1}$) \\
\hline

58030.18374	& 149.757 &	0.934 & 155.320 & 10 \\
58033.23587 & 149.333 &	0.983 & 154.478 & 10 \\
58055.11253	& 150.302 &	0.859 & 159.028 & 10 \\
58058.09523 & 150.309 &	0.775 & 157.339	& 10 \\
58061.15510 & 149.650 & 1.667 & 157.238	& 10 \\
58351.32561 & 158.755 &	0.926 & 125.104	& 10 \\
58356.28997 & 158.604 &	0.814 & 130.785	& 10 \\
58382.22408 & 159.618 &	0.928 & 131.047 & 10 \\
58386.27275 & 159.696 &	0.752 & 130.500 & 10 \\
58416.12653 & 159.215 &	0.700 & 123.162 & 10 \\ 
58442.07649 & 159.883 &	0.569 & 123.092 & 10 \\
58445.08565 & 160.084 &	0.757 & 111.369 & 10 \\
58682.38554 & 162.852 &	0.749 & 152.033 & 2.408 \\
58706.33070 & 164.036 & 0.552 & 154.207 & 1.347 \\
58709.35218 & 163.733 &	0.799 & 154.059 & 1.338 \\
58712.31703 & 163.844 &	0.874 & 154.894 & 1.234 \\
58766.18504 & 161.693 &	1.522 &155.860 & 3.204 \\
\hline
\end{tabular}
\tablecomments{This table is available in its entirety in machine-readable form.}
\end{table*}

\bibliography{biblio}{}
\bibliographystyle{aasjournal}

\listofchanges

\end{document}